\documentclass[twocolumn]{openjournal}
\usepackage{graphicx} % Required for inserting images

\usepackage[titletoc]{appendix}
\usepackage[utf8]{inputenc}
\DeclareUnicodeCharacter{03BC}{\ensuremath{\mu}}
\usepackage{xspace}
\usepackage[colorlinks=true,urlcolor=blue,citecolor=NavyBlue]{hyperref}
\setcitestyle{authoryear,open={(},close={)}} % openjournal
\usepackage{url}

\usepackage{tikz}

\makeatletter
\renewcommand\appendix{\par
  \setcounter{section}{0}%
  \setcounter{subsection}{0}%
  \renewcommand\thesection{\@Alph\c@section}%
  \renewcommand\thesubsection{\thesection.\@arabic\c@subsection}}
\makeatother

\def\referee#1{#1}
\def\reftwo#1{#1}

\usepackage{
	amsmath,
	amssymb,
	amsthm,
	microtype,
    stmaryrd,
    xcolor, 
}

\usepackage{placeins}
\usepackage[svgnames]{xcolor}

\providecommand{\apjl}{ApJL} % openjournal
\providecommand{\apjs}{ApJS} % openjournal
\providecommand{\aap}{A\&A} % openjournal
\renewcommand{\micron}{\ensuremath{\mu\mathrm{m}}\xspace}

\def\software#1{\textbf{Software:} \texttt{#1}\xspace} % openjournal
\newcommand{\kms}{\mathrm{km~s}^{-1}\xspace}

\newcommand{\orcidauthor}[3]{\author{\href{http://orcid.org/#1}{#2$^{#3}$}}}

\begin{document}

\title{Mapping CO Ice in a Star-Forming Filament in the 3 kpc Arm with JWST}

\orcidauthor{0000-0002-1313-429X}{Savannah R. Gramze}{1*}
% \email{savannahgramze@ufl.edu}
\orcidauthor{0000-0001-6431-9633}{Adam Ginsburg}{1}
% \email{adamginsburg@ufl.edu}
\orcidauthor{0000-0002-0533-8575}{Nazar Budaiev}{1}
% \email{nbudaiev@ufl.edu}
\orcidauthor{0000-0002-4407-885X}{Alyssa Bulatek}{1}
% \email{abulatek@ufl.edu}
\orcidauthor{0009-0001-8880-6951}{Theo Richardson}{1}
% \email{richardson.t@ufl.edu}
\orcidauthor{0000-0002-3941-0360}{Miriam G. Santa-Maria}{1,12}
% \email{miriam.g.sm@csic.es}

\orcidauthor{0000-0003-0410-4504}{A.~T.~Barnes}{2}
% \email{ashleybarnes.astro@gmail.com}
\orcidauthor{0000-0002-6379-7593}{Francisco Nogueras-Lara}{2,3}
% \email{fnoguera@eso.org}

\orcidauthor{0000-0001-6113-6241}{Mattia~C.~Sormani}{4}
% \email{mattiacarlo.sormani@gmail.com}
\orcidauthor{0000-0003-2619-9305}{Xing Lu}{5,6}
% \email{xinglv.nju@gmail.com}
\orcidauthor{0000-0003-4224-6829}{Brandt A. L. Gaches}{7}
% \email{brandt.gaches@uni-due.de}
\orcidauthor{0000-0002-6073-9320}{Cara D. Battersby}{8}
% \email{cara.battersby@uconn.edu}
\orcidauthor{0009-0002-7459-4174}{Jennifer Wallace}{8}
% \email{jennifer.2.wallace@uconn.edu}
\orcidauthor{0000-0001-7330-8856}{Daniel~L.~Walker}{9}
% \email{daniel.walker.astro@gmail.com}
\orcidauthor{0000-0001-8782-1992}{Elisabeth A.C. Mills}{10}
% \email{eacmills@ku.edu}
\orcidauthor{0009-0005-9197-6483}{Michael Mattern}{11}
% \email{michael.mattern@cea.fr}
\orcidauthor{0009-0004-0685-7678}{Rojita Buddhacharya}{13, 14}
% rojitabuddhacharya@gmail.com

%%%%%%%%%%%%%%%%%%%%%%%

\affiliation{$^{1}$Department of Astronomy, University of Florida, Gainesville, FL 32611 USA}
\affiliation{$^{2}$European Southern Observatory (ESO), Karl-Schwarzschild-Stra{\ss}e 2, 85748 Garching, Germany}
\affiliation{$^{3}$Instituto de Astrofísica de Andalucía, CSIC, Glorieta de la Astronomía s/n, E-18008 Granada, Spain}
\affiliation{$^{4}$Como Lake centre for AstroPhysics (CLAP), DiSAT, Universit{\`a} dell’Insubria, via Valleggio 11, 22100 Como, Italy}
\affiliation{$^{5}$Shanghai Astronomical Observatory, Chinese Academy of Sciences, 80 Nandan Road, Shanghai 200030, P.\ R.\ China}
\affiliation{$^{6}$State Key Laboratory of Radio Astronomy and Technology, A20 Datun Road, Chaoyang District, Beijing, 100101, P.\ R.\ China}
\affiliation{$^{7}$Faculty of Physics, University of Duisburg-Essen, Lotharstraße 1, 47057 Duisburg, Germany}
\affiliation{$^{8}$University of Connecticut, Department of Physics, 196A Auditorium Road, Unit 3046,
Storrs, CT 06269}
\affiliation{$^{9}$UK ALMA Regional Centre Node, Jodrell Bank Centre for Astrophysics, The University of Manchester, Manchester M13 9PL, UK}
\affiliation{$^{10}$Department of Physics and Astronomy, University of Kansas, 1251 Wescoe Hall Drive, Lawrence, KS 66045, USA}
\affiliation{$^{11}$Laboratoire d'Astrophysique (AIM)  Universit\'e Paris-Saclay, Universit\'e Paris Cit\'e, CEA, CNRS, AIM 91191 Gif-sur-Yvette, France}
\affiliation{$^{12}$Instituto de Física Fundamental (CSIC). Calle Serrano 121-123, 28006, Madrid, Spain}
\affiliation{$^{13}$Astrophysics Research Institute, Liverpool John Moores University, 146 Brownlow Hill, Liverpool L3 5RF, UK}
\affiliation{$^{14}$Center for Astrophysics | Harvard \& Smithsonian, 60 Garden Street, Cambridge, MA 02138, USA}

%%%%%%%%%%%%%%%%%%%%%%%%

%%% Blank 
%\author[]{}
%\affiliation{}
%\email{}

\begin{abstract}
CO gas emission is a fundamental tool for measuring column density, but in cold, dark clouds, much of the CO is locked away in ice. 
We present JWST results from observations of a star forming filament (G0.342+0.024) that appears to be associated with the 3 kpc arm. This filament is backlit by the \referee{Nuclear Stellar Disk}, which has allowed us to construct a high-resolution extinction map (mean separation between stars of $\sim1\arcsec$ outside the filament, $\sim2 \arcsec$ in the filament). ALMA Band 3 data reveals embedded star formation within the cloud. 
Using the CO ice feature covered by the F466N band, we map the CO ice column density of the filament. 
By combining the extinction map, CO ice column density map, and archival CO observations, we examine the efficacy of standard CO X-factor measurements of mass in star forming gas. 
We find that 50-88\% of the CO is locked away in ice at large column densities ($N_{\mathrm{H_2}}\gtrsim~\mathrm{1\times10^{22}~cm^{-2}}$, $200~\mathrm{M_\odot~pc^{-2}}$) in the filament. 
The primary sources of uncertainty in this estimate are due to uncertainty in the ice composition and lab measurements of ice opacities. 
\referee{
We measure CO column densities that surpass the limits of locally-measured CO abundances, suggesting a metallicity-dependent Galactic variation in CO abundance. 
This measurement implies that systematic corrections are needed for mass measurements in high column density regions of the ISM. 
}
\end{abstract}

\section{Introduction}

\referee{
CO is used to trace mass in galaxies and molecular clouds, usually via the emission lines it produces in the gas phase. 
While spectral CO gas measurements allow us to select for only the relevant gas via velocity component association, 
``CO dark" gas can cause mass underestimates \citep{Peretto2023P}, while large linewidths \citep{gramze2023} and metallicity differences \citep{Bolatto2013, Gong2020, Bisbas2025, Kohno2024} can make measurements with the CO X-factor \citep{strong88} inaccurate.
}

\referee{
As CO molecules freeze out of the gas phase, they no longer contribute to CO gas emission, throwing off mass measurements of clouds that use it
\citep{Hernandez2011, Gainey2022, clarke2024, Cosentino2025}.
Infrared dark clouds (IRDCs) are regions of the interstellar medium (ISM) where this happens, as they are cold \citep[T $\leq20~\mathrm{K}$][]{Pillai2006} and dense \citep[$\mathrm{n_H} \geq 1\times10^4~\mathrm{cm^{-3}}$][]{Butler2012} enough for catastrophic freeze-out of CO \citep{Caselli2022}. 
}

\referee{
IRDCs have not yet been investigated in the same level of depth for their ice composition as local molecular clouds. 
These clouds have been surveyed with JWST to reveal a rich mixture of different ice species imprinted on the ISM \citep{McClure2023, Rocha2024, smith_cospatial_2025}. 
Prior to JWST, interstellar ices have mainly been studied using spectra from ISO \citep{Gibb2004}, Spitzer \citep{Boogert2008, Oberg2008, Pontoppidan2008}, and some ground based observatories \citep{Jang2022}. 
While SPHEREx will soon provide spectroscopic measurements toward millions of stars backlighting thousands of IRDCs, its poor spatial resolution will limit interpretation, especially toward the Galactic Center.
JWST's superior resolution and sensitivity now allow us to measure ices within IRDCs.
}

\referee{
In this work, we utilize JWST observations of the Central Molecular Zone (CMZ) of the Galactic Center.
The Galactic Center has one of the highest stellar densities of any part of the sky \citep{Sormani2020jeans}, 
%(the total mass of the nuclear stellar disk at the center of the Galaxy is $6.9 \pm 2 \times 10^8 ~\mathrm{M}_\odot$ \citep{Sormani2020jeans}), 
which is visible in infrared observations that peer through the dust in front of the Galactic Center \citep{Churchwell2009, Nogueras2019, NoguerasLara2021b}. 
IRDCs in front of the Galactic Center are backlit by thousands of stars that are reddened by the dense, dusty interiors of the clouds. 
}

\referee{
Interposed between the Sun and the Galactic Center are several spiral arms: Sagittarius–Carina, Scutum–Centaurus–OSC, Norma–Outer, and the $3~\mathrm{kpc}$ arm \citep{reid2019, NoguerasLara2021a}. 
These spiral arms encompass much of the stars, gas, and dust sitting in front of the Galactic Center. 
Serendipitously, within our observations, we observe an IRDC filamentary cloud in front (justified in Section \ref{sec:distance}) of the Galactic Center.
}

\referee{
In this paper, we present observations and analysis of the properties of a star-forming filament in front of the Galactic Center, backlit by thousands of stars. 
In Section \ref{sec:obs}, we provide an overview of the observations used to analyze the filament. 
Next, in Section \ref{sec:datareduction}, we explain our reduction of the data.
Then in Section \ref{sec:analysis}, we describe the data qualitatively and measure the dust extinction and CO ice column density toward each star.
Section \ref{sec:results} is dedicated to presenting the results of our analysis. 
Finally, in Section \ref{sec:discussion}, we demonstrate that the filament is in front of the Galactic Center and has an overabundance of CO compared to local measurements. 
}

\section{Observations} \label{sec:obs}

\begin{figure*}
    % lactea-filament/notebooks/big_picture.ipynb
    \centering
    \includegraphics[width=\linewidth]{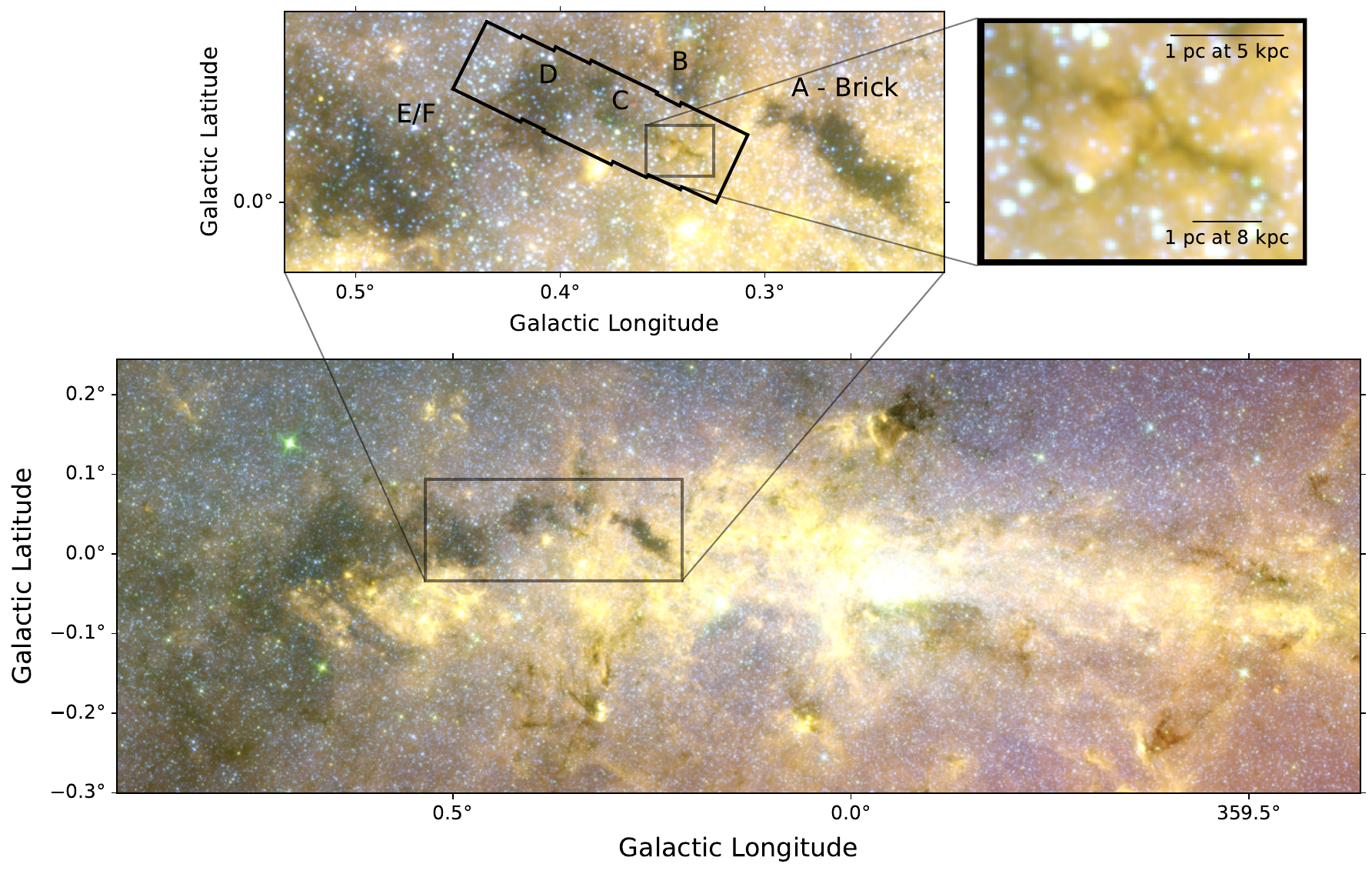}
    \caption{Overview figure using Spitzer I4, I3, and I1 \citep{Churchwell2009}. The bottom panel shows the context of the Galactic Center Central Molecular Zone. The top left panel zooms in on the dust ridge region, labeling each of the dust ridge clouds \citep{lis1999}, and showing the field of view of the JWST NIRCam \referee{field of view} (the jagged, zig-zag field). The top right panel shows a zoom in of the Filament.}
    \label{fig:overview}
\end{figure*}

We observe the Galactic Center dust ridge with JWST. 
These observations include IRDC dust ridge clouds C (G0.380+0.050) and D (G0.412+0.052) and a foreground IRDC filament (G0.342+0.024).
These NIRCam observations were taken in parallel as a bonus to concurrent MIRI observations of the Brick, an IRDC with low star formation thought to be located in the Galactic Center, as part of JWST GO Program 2221.
The footprint of these NIRCam observations in relation to the dust ridge is shown in Figure \ref{fig:overview} overlaid on a Spitzer 3-color image of IRAC $8$, $5.8$, and $3.6~\micron$ bands. 
% 3.6, 4.5, 5.8 and 8.0 µm
The irregular shape of the footprint is due to the observations running parallel to the MIRI observations of the Brick, so the field does not have the standard rectangular NIRCam dither pattern.
The observational setup is the same as observations of the Brick in the same program, described in \citet{ginsburg2023}. 

For this program, we use 6 NIRCam imaging filters (3 each in the short wavelength and long wavelength channels). 
JWST NIRCam achieves $0.031\arcsec~\mathrm{px^{-1}}$ 
resolution for short wavelength and $0.063\arcsec~\mathrm{px^{-1}}$ for long wavelength filters. 
These filters were chosen to cover hydrogen recombination lines (F405N and F187N), their continuum counterparts (F410M and F182M), molecular $\mathrm{H_2}$ emission (F212N), and CO emission (F466N). 

We utilize continuum and molecular line emission observations from the Atacama Large Millimeter/submillimeter Array (ALMA) large program ALMA Central Molecular Zone Exploration Survey (ACES) (project id: 2021.1.00172.L) in Band 3 
\referee{\citep{Longmore2026, Ginsburg2026, Walker2026, Lu2026, Hsieh2026}}.
%(Longmore et al in prep, Ginsburg et al in prep, Walker et al in prep, Lu et al in prep, Hsieh et al in prep). 

Additionally, we use single dish $\mathrm{^{12}CO}~J=1\rightarrow0$,  $\mathrm{^{13}CO}~J=1\rightarrow0$ and $\mathrm{C^{18}O}~J=1\rightarrow0$ observations from the $\mathrm{45~m}$ Nobeyama Radio Observatory (NRO) telescope BEARS and FOREST receivers \citep{Tokuyama2019}. 

\section{Data Reduction}
\label{sec:datareduction}

We use a similar pipeline to \cite{ginsburg2023} to conduct data reduction of the JWST observations presented in this paper. First, we download the data from MAST (Barbara A. Mikulski Archive for Space Telescopes) via \texttt{astroquery}. 
\referee{The data described here may be obtained from
\url{https://dx.doi.org/10.17909/zpm2-2258}.}
We receive 32 exposures per filter taken in a total of two visits. Next, we reprocess the images starting from the \texttt{\_uncal} files. For Stage 1 of the pipeline, we set \texttt{suppress\_one\_group} to \texttt{False} and \texttt{use\_side\_ref\_pixels} in \texttt{refpix} to \texttt{True}. 

\subsection{Destreak} \label{sec:destreak}

\begin{figure}
    % pink_noise/notebooks/compare_residuals.ipynb
    \centering
    \includegraphics[width=\linewidth]{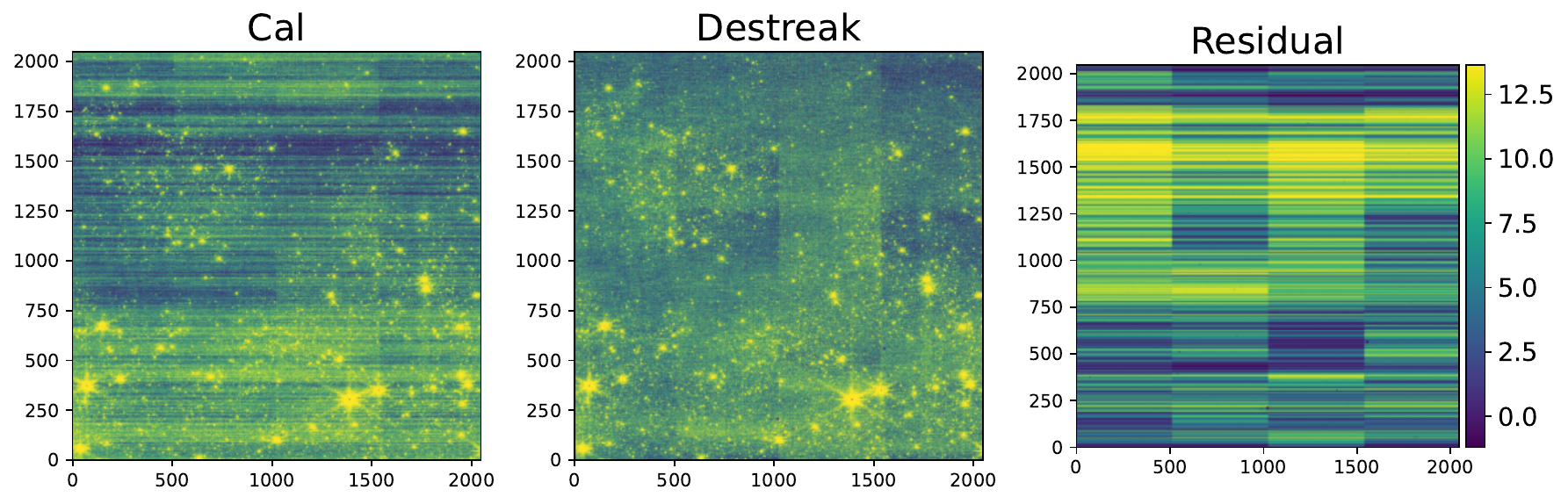}
    \caption{The process of "destreaking" the \texttt{\_cal} images. Left is the starting \texttt{\_cal} image. Middle is the image after 1/f noise reduction. Right is the residual, or difference between the two images. The process is not perfect, as \referee{seen} by the difference in the columns/amplifiers in the middle image.}
    \label{fig:destreak}
\end{figure}

We find that our short wavelength filter observations are affected by 1/f noise, which leaves ``streaks" across the calibration images. 
This effect is especially prominent in the low signal regions of the images, where the molecular clouds are located. 
We use an adjusted method to the one described in \citet{ginsburg2023} to attempt to remove 1/f noise from the images. 

%Between stages 2 and 3 of the pipeline, 
Before the pipeline-calibrated images are made into science-ready mosaics, 
we perform a ``destreak" step based on a method by Massimo Robberto (private communications) to mitigate 1/f noise. 
First, each \texttt{\_cal} image is split into four horizontal quadrants with a width of $\mathrm{512~px}$ and a height of $\mathrm{2048~px}$. 
Next, we calculate the median across the horizontal axis, resulting in a $\mathrm{2048~px}$ median array.
We find that the destreak method acts like a high pass filter at this stage, removing significant amounts of extended emission, so we obtain images retaining the large scale structure of the field by first running the pipeline without any 1/f noise removal techniques. 
From these images, we compute percentile filter background images to estimate the ``zero spacing" background level across the images. 
The resulting images are shown in Appendix Figure \ref{fig:app-perfilt}. 
We make these images by applying a \texttt{scipy.ndimage.percentile\_filter} with a circular footprint ($\mathrm{64~px}$ for F410M, F405N, F466N and $\mathrm{128~px}$ for F182M, F187N, F212N) and then taking the 10th percentile of the result to get the percentile filter background image. 
We then smooth the median array by the percentile field background. 
Finally, we subtract the median array from each quadrant, then add back the smoothed median array. 

\referee{We evaluate the effectiveness of this process by comparing the output of the 1/f noise mitigation method with the initial input.
Figure \ref{fig:destreak} shows a comparison between the initial \texttt{\_cal} file, its destreaked version, and the residual.  
Areas with bright emission in the data, such as where there are bright stars or areas of diffuse emission, \textbf{poorly} recovered, resulting in artifacts like horizontal smears.
While the 1/f noise is mitigated in the short wavelength filters, some streaking remains.
Other methods of 1/f noise removal were tested\footnote{https://github.com/SpacialTree/pink\_noise/blob/master/\\notebooks/compare\_residuals.ipynb} and compared to our method, but ``destreak" removes 1/f noise the most effectively in this case where the data have both high stellar density and extended structure. A significant fraction of the 1/f noise is removed, while the large scale structure is largely retained.}

\subsection{Image Alignment} \label{sec:alignment}

These data are impacted by a known issue with the JWST Fine Guidance Sensor (FGS), which since JWST Cycle 3 has now been mitigated by adding the GALACTICNUCLEUS survey \citep{Nogueras2019} to the guide star catalog.
Occasionally, especially in crowded fields such as the Galactic Center, the FGS will lock onto the wrong star, resulting in a loss of positional accuracy. 
The FGS can also lock onto different stars per visit. 
This error in pointing then propagates to the reported NIRCam pointing, resulting in offsets of several arcseconds and poorly-aligned mosaics. 
We mitigate this problem with our data by first measuring the positional offset by comparing the locations of the stars with those from the Vista Variables in the Via Lactea (VVV) survey catalog \citep{Smith2025}. 
Visit 001 is offset by $\sim8\arcsec$, and Visit 002 is offset by $\sim4\arcsec$. We then use the \texttt{tweakreg} utility function \texttt{adjust\_wcs} to shift the Generalized World Coordinate System (GWCS) and FITS WCS of each \texttt{\_cal} image before running it through \texttt{tweakreg}, so that the function has a better initial guess for the alignment of the images. 

Further offsets noted in \citet{ginsburg2023} between the two NIRCam modules were mitigated by the February 15, 2024 update to the JWST CRDS files. 

We use the VVV catalog \citep{Smith2025} as an initial reference catalog instead of the default Gaia catalog. There are far fewer stars overlapping between the Gaia and JWST catalogs than with VVV, and the Gaia stars in our field are often oversaturated. We first use the VVV catalog as a reference catalog for F405N and then make a new reference catalog from F405N. We detail the creation of the F405N catalog in Section \ref{sec:cataloging}. For all other filters, we use the F405N catalog as the reference for image alignment.

\subsection{Cataloguing} \label{sec:cataloging}
After correcting for image alignment, we begin cataloging the data using the now properly-aligned individual frames. 
First, we begin by modeling the Point Spread Function (PSF). 
We utilize the \texttt{webbpsf} package, a model based on tracing the optical path of JWST and taking the fourier transform of the effective aperture. 
We make a grid of PSF models for each filter. 
We use the oversampled PSF with distortion corrections (OVERDIST), making a grid of models across the field. 
We use the OVERDIST PSF model as it adds less numerical noise or uncertainty. 

We then use \texttt{photutils} \citep{larry_bradley_2025} to perform PSF photometry on each \referee{frame}.
We use the post-outlier correction \texttt{\_crf} files for our photometry, as they go through the 1/f noise correction step detailed in Section \ref{sec:destreak}.  
We calculate \referee{Vega} magnitudes by acquiring zero points from the SVO Filter Profile Service \citep{Rodrigo2024}, and we then use these magnitudes throughout this paper.

After cataloging each frame, we then combine each frame's catalog into a single catalog per filter.
Finally, we combine each filter's catalog into one single catalog for the entire field, which ensures that stars detected in multiple filter catalogs are combined into a single row. 
We set the minimum offset to create a new row for a star to $0.01\arcsec$, meaning that any stars below that separation are assumed to be the same star, as the angular resolution of JWST is $\sim0.1\arcsec$ with the pixelscale of the NIRCam detector being $0.031\arcsec~\mathrm{px^{-1}}$ for short wavelengths and $0.063\arcsec~\mathrm{px^{-1}}$ for long.

Diffraction spikes are often erroneously cataloged. To mitigate the inclusion of these false positive detections, we mask the catalog of stars to try and remove them. First, we include only stars with a quality factor $\texttt{qf < 0.4}$. The cataloging algorithm calculates magnitude errors, so next we select for stars with magnitude errors $\texttt{emag < 0.1}$. Then we remove stars that are detected in only one band, requiring a detection in at least two. Lastly, there is a population of bright, real stars that are categorically improperly fit. We remove these via a color and magnitude cut, masking out stars with [F405N] - [F410M] $< -0.3$, [F405N] $< 13.5$ excluding non-detections, [F187N]-[F182M]$<-0.3$, and [F187N]$<15$ excluding non-detections. The resulting final catalog is more accurate at the cost of completeness. 

\section{Analysis} \label{sec:analysis}

\subsection{Target Description}
\label{sec:targetdescription}

\begin{figure*}[!ht]
    % lactea-filament/notebooks/three-color-cloudc.ipynb
    \includegraphics[width=\textwidth]{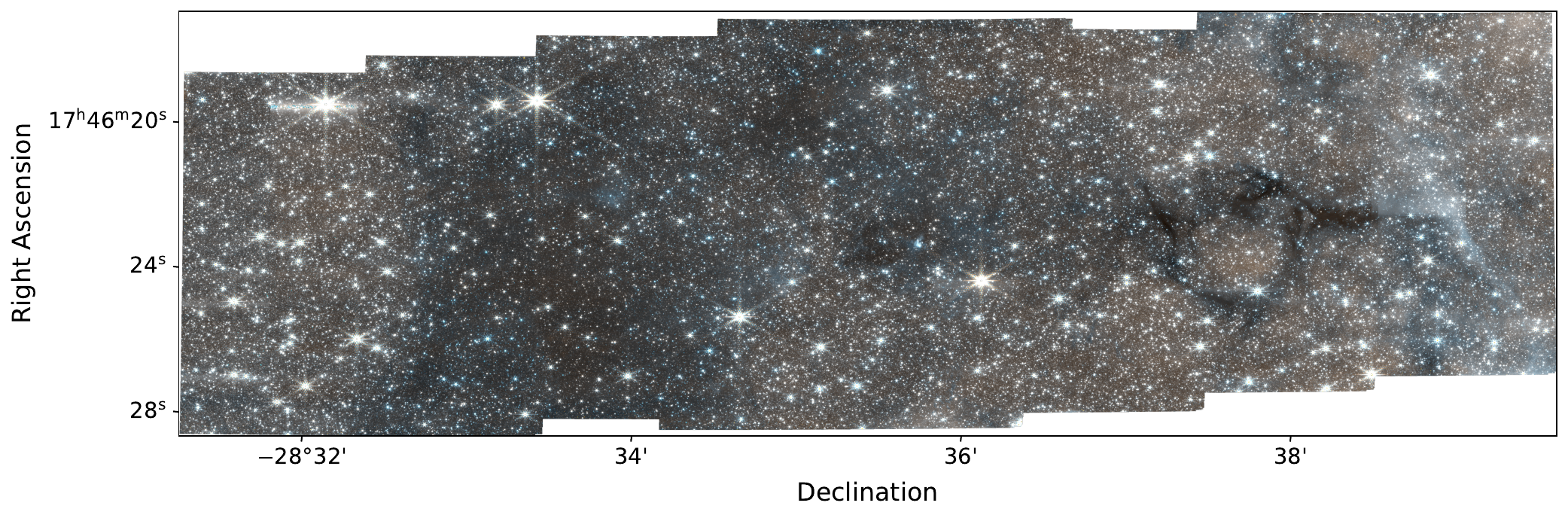}
    \caption{Three color image of the region observed with JWST in parallel. Red is F466N, blue is F405N, and green is their combination. Several IRDCs are visible in this image.  \referee{Appendix Figure \ref{fig:cropped} is a version of this figure with each of the distinct IRDCs labeled.}}
    \label{fig:jwst-fov}
\end{figure*}

In Figure \ref{fig:overview}, we show the Galactic Center, a zoom-in panel of the ``dust ridge" with each cloud labeled and the outline of the JWST \referee{field of view} shown, and a final panel showing a filamentary structure in extinction. In this paper, we refer to this feature as the filament.
The filament has a projected size $\sim2\arcmin$ across the field. 

The imaging results of the data reduction pipeline are shown in a three color image in Figure \ref{fig:jwst-fov}. 
On the right side of the image is the dark filamentary structure associated with the filament in Figure \ref{fig:overview}. The filament appears very structured, with elongated structures that seem to represent dense material and gaps where there is less material. 

In the middle and left of the image are Galactic Center dust ridge clouds C and D respectively. Figure \ref{fig:overview} labels the clouds in the top left panel. 
At the top right of the image is a dark feature likely associated with dust ridge cloud B. 
A diffuse HII region runs vertically on the right side of Figure \ref{fig:jwst-fov}, which is most likely behind the filament and not associated with foreground gas. 

\begin{figure*}[!ht]
    % lactea-filament/notebooks/filament_ccd.ipynb
    \centering
    \includegraphics[width=0.9\linewidth]{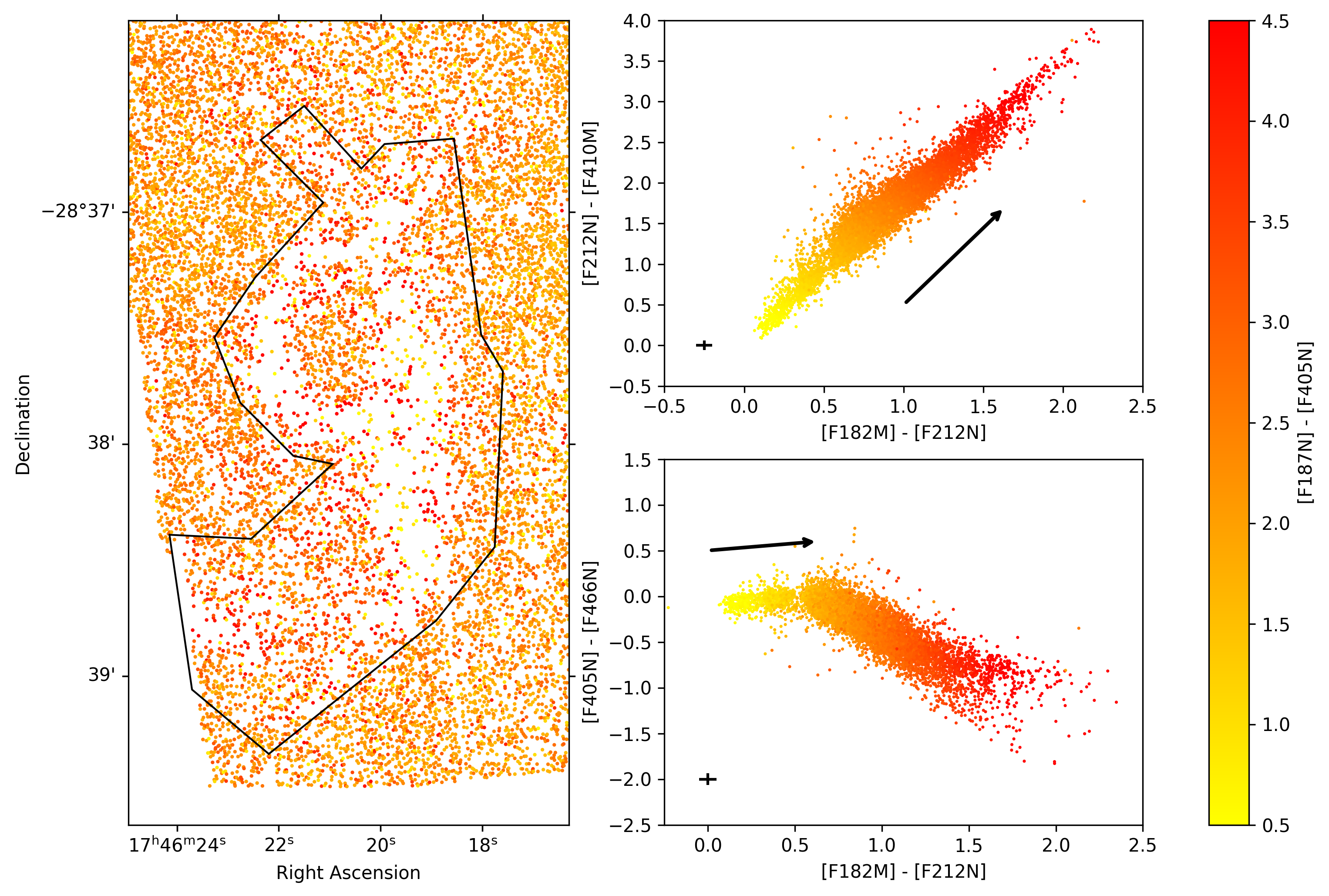}
    \caption{The left panel shows the locations of $37344$ stars detected towards the filament with the color map showing the color changes due to dust extinction in [F187N]-[F405N]. The region used for close-in analysis of the filament is over-plotted. The top right panel shows how all three colors [F182M]-[F212N], [F212N]-[F410M], and [F187N]-[F405N] are reddened by dust extinction. The bottom right panel shows how [F405N] - [F466N] differs from the other colors, becoming bluer with increasing extinction past a certain point, due to the inclusion of [F466N], which is affected by CO ice in the band. The black vectors in the color-color diagrams show the direction of reddening from dust extinction, and it shows how much a star at their origin would be reddened or moved in color-color space by $\mathrm{A_V}=20$ of extinction using the \texttt{CT06} extinction law. \referee{The black marker in the bottom left of each CCD represents the typical errorbars for each point, each on the order of $\mathrm{0.05~mag}$.}} 
    \label{fig:ccd}
\end{figure*}

\subsection{Color-Color Diagrams} \label{sec:ccd}
Using the catalog created in Section \ref{sec:cataloging}, we create Figure \ref{fig:ccd}. The left panel shows the positions of stars included in the catalog colored by [F187N]-[F405N]. Notably, the stars become redder in this color the closer they are to the interior of the filament, which is distinguished by the smaller stellar density along its length. This effect is due to extinction, where high column densities of dust block out light from background stars, especially at bluer wavelengths. While most of the stars near the filament appear reddened, there is also a foreground population of stars. 

Dust extinction affects shorter wavelengths more strongly than longer, so the color of two filters is proportional to the extinction. 
The right panels of Figure \ref{fig:ccd} show two color-color diagrams (CCDs). The top right panel shows how all three colors [F182M]-[F212N], [F212N]-[F410M], and [F187N]-[F405N] 
are influenced primarily by dust extinction, with all three colors increasing linearly to the right.

The bottom right panel of Figure \ref{fig:ccd} shows [F405N]-[F466N] on the vertical axis. Instead of increasing linearly as with the other colors, [F405N]-[F466N] trends downward after [F182M]-[F212N]$ > 0.55$, meaning that dust extinction does not dominate past this color. 
F466N overlaps with a strong CO ice absorption feature at $4.673~\micron$ \citep{ginsburg2023}. The presence of CO ice along the line of sight to highly extincted stars causes their light to be absorbed, making the [F466N] magnitudes of the stars dimmer and [F405N]-[F466N] seem bluer as a result.  
The \citet{fritz2011} extinction law takes into account the effects of ice absorption, but not enough to account for the ice absorption observed in the filament. 

\reftwo{While F466N is the reddest filter used in this survey, we expect that the CO ice feature it traces is narrow, and the stellar continuum recovers at logner wavelengths past it. Figure 22 of Appendix J in \citet{Ginsburg2025} shows ISO spectra of stars associated with the Quintuplet cluster, also taken toward the Galactic Center, which shows that the continuum of the stars recovers past the CO ice feature. Spectra from \citet{smith_cospatial_2025} of nearby low mass star forming region Chameleon I also show that the continua of background stars recover after the CO ice feature.}

\subsection{Extinction Map} \label{sec:extinction}

\begin{figure}
    \centering
    \includegraphics[width=\linewidth]{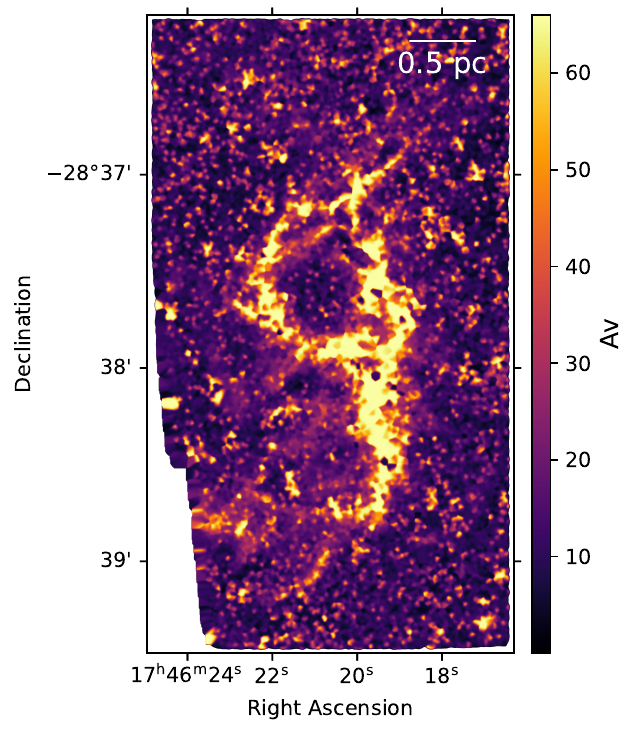}
    \caption{Extinction map \referee{of} the filament created using $\mathrm{A_V}$ measured with [F182M]-[F410M]. Stars behind the filament that were detected in F410M but not F182M were given a flat $A_V=85$, then a constant value of $\mathrm{22~mag}$ was subtracted from the whole map to remove foreground extinction.}
    \label{fig:extinction_map}
\end{figure}

We create an extinction map of the filament using [F182M] - [F410M] as our reference color. 
%We create an extinction map of the filament. 
First, we select a region around the filament to analyze using \texttt{regions} \citep{larry_bradley_2024_13852178}, including only stars within the region immediately around the filament to analyze. 
Using the catalog made in Section \ref{sec:cataloging}, we select only stars toward the filament region. We remove stars from the catalog with [F182M]-[F212N] $< 0.55$ and [F182M]-[F410M] $< 2$, ensuring we select only stars inside or behind the filament. This removes foreground contaminants in the measured extinction. We measure the extinction toward each star using: 

\begin{equation} \label{eq:extinction}
    %av182410 = (self.color('f182m', 'f410m')) / (ext(1.82*u.um) - ext(4.10*u.um))
    A_V = \frac{[F182M] - [F410M]}{ext(1.82~\mu m) - ext(4.10~\mu m)}
\end{equation}

\noindent
where the difference in the magnitudes of stars in the F182M and F410M bands is divided by the difference in the extinction (\emph{ext}) law at the representative wavelength of the bands, assuming the \texttt{CT06} extinction law \citep{chiar2006} using the \texttt{dust\_extinction} python package \citep{Gordon2024} (see Section \ref{sec:ext_mass} below \reftwo{and Appendix E2 of \citet{Ginsburg2025}} for more discussion about the impacts of different extinction laws). 
For stars detected in F410M without a detection in F182M, we assume that the extinction toward that star is $A_V=85$, the maximum extinction measured with [F182M]-[F410M] with \texttt{CT06}. 

Next, we create a blank grid the size of the region with a pixel resolution of $\mathrm{0.03~\arcsec~pix^{-1}}$, matching the resolution of the short wavelength images. 
At the pixel coordinates of each cataloged star, we set the grid value to the measured extinction toward the star. 
For pixels in the grid that correspond to gaps in the short wavelength NIRCam filter mosaics, where stars are detected in the long wavelength filters, \referee{we reset the pixel values in the grid to blank to avoid overestimating the extinction and to prevent linear artifacts at the gaps.}

Then, we use the \texttt{astropy} convolution method \texttt{convolve\_fft} and a 2D Gaussian with a FWHM of 15 pixels and a kernel size of 181 to interpolate over the grid and fill it in with the measured extinction values. This 2D kernel does the best job of filling in the low stellar density center of the filament while retaining as much of the resolution in the rest of the map, where the stellar density is higher, as possible.  

To remove foreground extinction from the map, we subtract a constant foreground extinction value of $A_V([\mathrm{F182M}]-[\mathrm{F410M}]=2)=\mathrm{21.69~mag}$ from the map, matching the color cut we made to select stars impacted by extinction from the filament. 
\citet{NoguerasLara2021a} uses $A_V(\mathrm{H}-\mathrm{K_s} = 0.9)=13~\mathrm{mag}$ for the fourth spiral arm from the Sun in front of the Galactic Center, but using this lower cut increases the amount of contaminant stars and increases the dust extinction. 

The resulting extinction map is shown in Figure \ref{fig:extinction_map}. The center of the filament likely reaches higher extinction values than the assumed maximum of $A_V=63$ that replaces stars that were not detected in F182M but were in F410M. The map likely underestimates the extinction in these areas, especially where the stellar density is lowest at the very center of the filament where detections are lacking even in F410M. Some areas of low extinction in the center of the filament are likely due to stars inside of the filament but not behind the majority of its dust. 

The extinction map assumes the \citet{chiar2006} extinction law. 
\referee{While this extinction law is measured toward the CMZ, the extinction curve likely varies for different sight lines  \citep{NoguerasLara2019}.}
We show extinction vectors in Figure \ref{fig:ccd} that predict the shift of a star in color-color space behind $20~\mathrm{mag}$ of extinction using the \texttt{CT06} extinction law. The vectors seem to parallel the observed distribution of stars, excluding colors impacted by CO ice absorption. 
% That said, other extinction laws differ by as much as 
Additionally, the extinction map does not take into account the intrinsic colors of the stars, and we assume that any reddening is due to extinction. 

\subsection{CO Ice} \label{sec:co_ice}

\begin{figure}
    \centering
    \includegraphics[width=\linewidth]{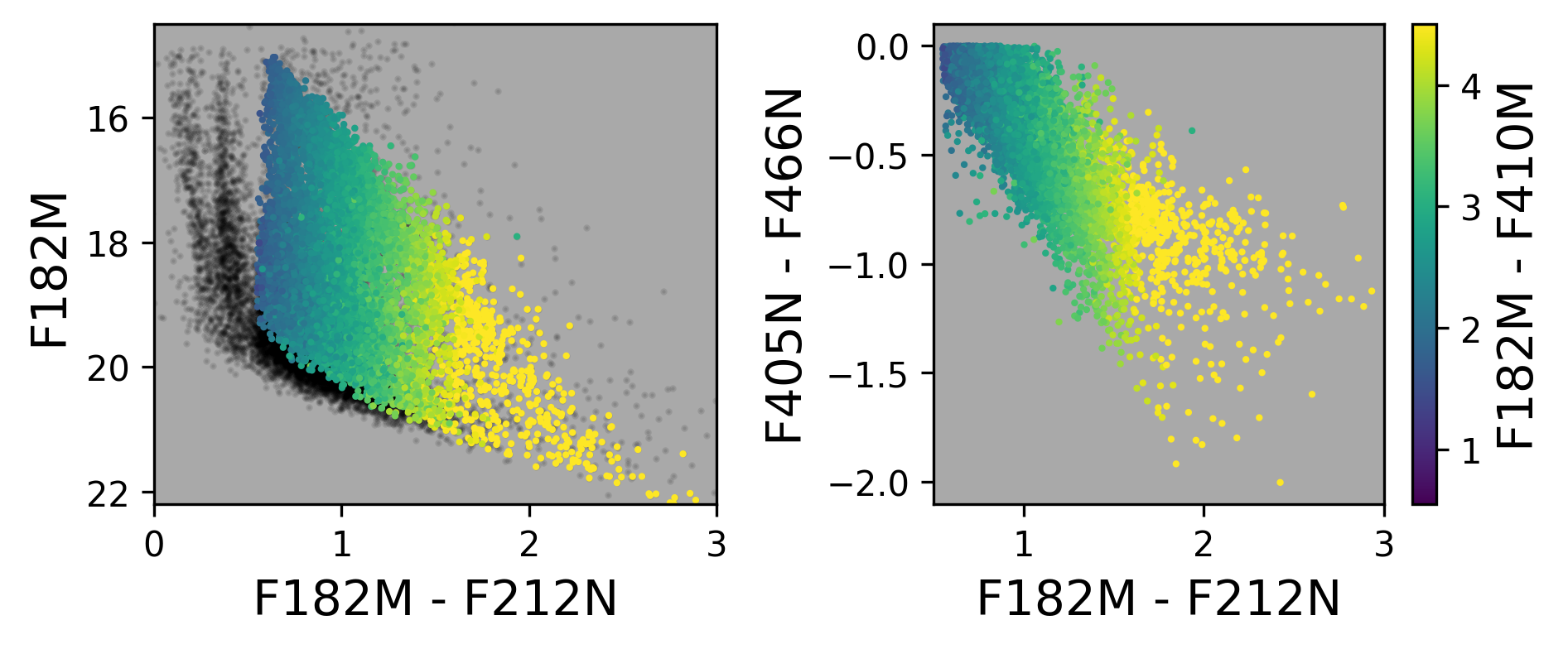}
    \caption{
    Left: Color-magnitude diagram showing the selection of sources used to create the CO ice column density map.
    Right: Color-color diagram of the selected sources, which have [F405N]-[F466N]$<0$.
    }
    \label{fig:COice_ccd}
\end{figure}

\begin{figure*}
    % CO_Extinction_filament.ipynb
    \centering
    \includegraphics[width=0.7\linewidth]{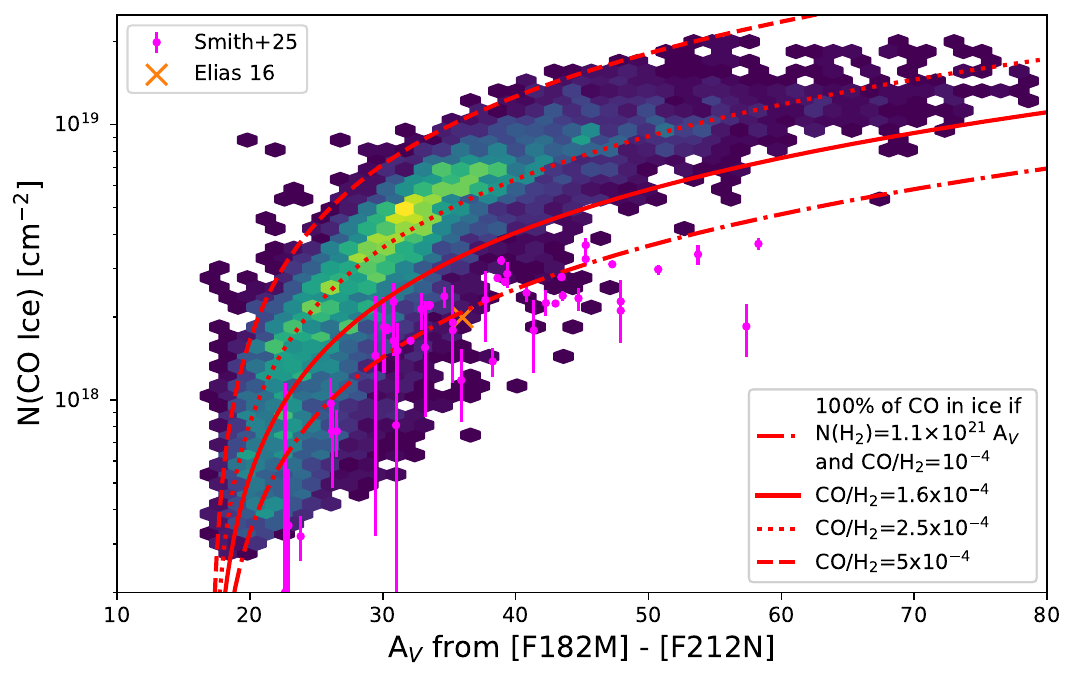}
    \caption{
    \referee{The measured extinction versus column density of CO ice for each star using [F405N] - [F466N]. 
    The magenta dots and error bars represent CO ice column densities measured by \citet{smith_cospatial_2025} with spectroscopy, which includes stars from \citet{McClure2023}, artificially offset by $17~\mathrm{mag}$ to overlap with the filament. 
    The orange X is from \citet{Knez2005}, also shifted to the right by $17~\mathrm{mag}$. 
    The red lines represent the theoretical limit of how much CO can be in the solid phase for different abundances of CO, where 100\% of the CO is frozen out of the gas phase, assuming the $\mathrm{H_2}$$/\mathrm{A_V}$ from \citet{Guver2009}.
    The solid red line uses the \referee{local CO abundance} from \citet{Lacy2017}.}}
    \label{fig:COice_Av}
\end{figure*}

\begin{figure}
    \centering
    \includegraphics[width=\linewidth]{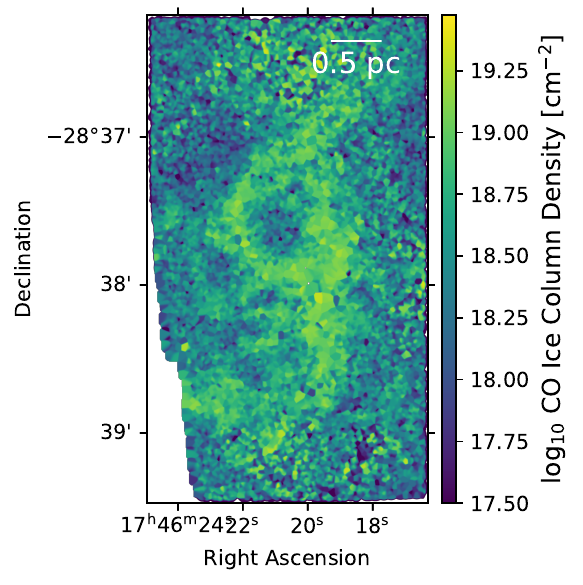}
    \caption{Map of CO ice column density over the filament made using [F405N] - [F466N] and modeled magnitude differences due to CO ice, using a $\mathrm{H_2O}$:CO:$\mathrm{CO_2}$ ratio of 10:1:1.}
    \label{fig:COice_map}
\end{figure}

We create a map of CO ice column density N(CO) toward the filament.
Motivated by the impact of CO ice on the F466N filter noted in Section \ref{sec:ccd}, we use this filter to estimate the column density of CO ice along the line of sight by modeling how much the stars are dimmed by the CO ice feature. 
Ice mapping has been done before using spectroscopic targets with JWST NIRCam's wide field slitless spectroscopy \citep{Smithice2025}. 
This new method takes advantage of NIRCam imaging to map the distribution of CO ice in a field with a high stellar density with thousands of targets.

First, we select stars within the region of the filament. We filter the stars to select only those with [F182M]-[F212N]$>0.55$, the same color limit used to make the extinction map, and [F405N]-[F466N]$<0$, which are reddened stars that are substantially extincted by dust to be either inside of or behind the filament (or even further behind in the Galactic Center), and stars that have been affected by CO ice enough to dim [F466N] below [F405N]. Figure \ref{fig:COice_ccd} shows the selected stars in color-magnitude and color-color space. 

In order to measure column density of CO ice, we model the effect of CO ice on the brightness of the stars using the \texttt{icemodels} package\footnote{https://github.com/keflavich/icemodels/}. 
Next, we combine laboratory measurements of the opacity profiles of CO \citep{Gerakines2023}, $\mathrm{H_2O}$ \citep{Mastrapa2009}, and $\mathrm{CO_2}$ \citep{Gerakines2020}, all of which significantly impact the filters used by this analysis. 
The $\mathrm{OCN^-}$ ice feature also overlaps with the F466N filter's wavelength coverage, but was deemed an insignificant contribution. 
We use a $\mathrm{H_2O}$:CO:$\mathrm{CO_2}$ ratio of 10:1:1, assuming a similar ice composition to the \citet{McClure2023} ice inventory toward background star J110621. 
With the combined ice mix opacities, we then model the impact of different column densities of the mixture on the light from a $\mathrm{4000~K}$ star using stellar atmospheric models \citep{Kurucz1979} \footnote{Variations in stellar atmospheres in the observed filters impact magnitudes by \referee{$< 0.1~\mathrm{mag}$} for stars with temperatures hotter than $\mathrm{4000~K}$. Intrinsic stellar colors do not produce systematic effects in the ice absorption bands, though they do have some effect in the short wavelength bands (\referee{$<2~\micron$}).}. 
We then find the magnitudes of the modeled star with and without modeled CO ice absorption, and the differences between these magnitudes. 
Afterward, we readied our measurements by removing the extinction from [F466N]-[F405N] using extinction measured with [F182M]-[F212N] \reftwo{as measured in Section \ref{sec:extinction}.}
Finally, we used the calculated relationship between [F466N]-[F405N] and N(CO) to measure the column density for each star. 

The results of measuring the CO ice column density toward each star within the constraints are shown in Figure \ref{fig:COice_Av}, which shows extinction $\mathrm{A_V}$ versus N(CO Ice). 
\referee{\citet{smith_cospatial_2025} measured the CO ice column density toward stars in the Chameleon I molecular cloud, $192~\mathrm{pc}$ from the Sun, which we have plotted in Figure \ref{fig:COice_Av}. These data points have been shifted horizontally by $\mathrm{17~mag}$ in order to overlap with the filament measurements.
The red lines shows the supposed maximum N(CO) at each $\mathrm{A_V}$ for if $100\%$ of the CO was in the ice phase at various CO/$\mathrm{H_2}$ abundances, assuming the \citet{Guver2009} relationship between $\mathrm{H_2}$ and $\mathrm{A_V}$.}

To map the spatial distribution of CO ice, we placed the measured column densities on a grid with their positions relative to each star, and then interpolated over the grid by convolving it with a 2D Gaussian kernel with a FWHM of 15 pixels, where the pixel resolution is $0.03~\arcsec~\mathrm{px}^{-1}$. The resulting map is shown in Figure \ref{fig:COice_map}, which shows the column density of CO ice. 
As a caveat, if stars are not detected, or are only detected in only one band, they cannot contribute to the CO ice map, so the ice column is a lower limit in the densest regions where the stellar density is low. 

The CO column density is highest in regions corresponding to the filament, morphologically matching its shape. 

\subsubsection{Systematic uncertainty on ice column density} \label{sec:ice_uncertain}
Our measurements of the CO ice column density rely on the adopted extinction curve, the relative mixture of ices present, the laboratory-measured opacities, and the input stellar models, roughly in that order.
The uncertainties driven by ices are examined in more detail in \referee{\citet{Ginsburg2025}}.
The uncertainty produced by \referee{the} adopted extinction curve is probably dominant, as \referee{using different extinction laws changes the measured by a factor of two (see Section \ref{sec:fil_mass} and Table \ref{tbl:mass}).} 
\reftwo{A more thorough exploration of the effects of using difference extinction laws and their effects on our measurements is in companion paper \citet{Ginsburg2025}.}

\section{Results} \label{sec:results}

\subsection{Velocity of the filament}

\begin{figure}
    % /home/savannahgramze/orange_link/adamginsburg/jwst/cloudc/notebooks/cloudc_spectral_fitting.ipynb
    \centering
    \includegraphics[width=\linewidth]{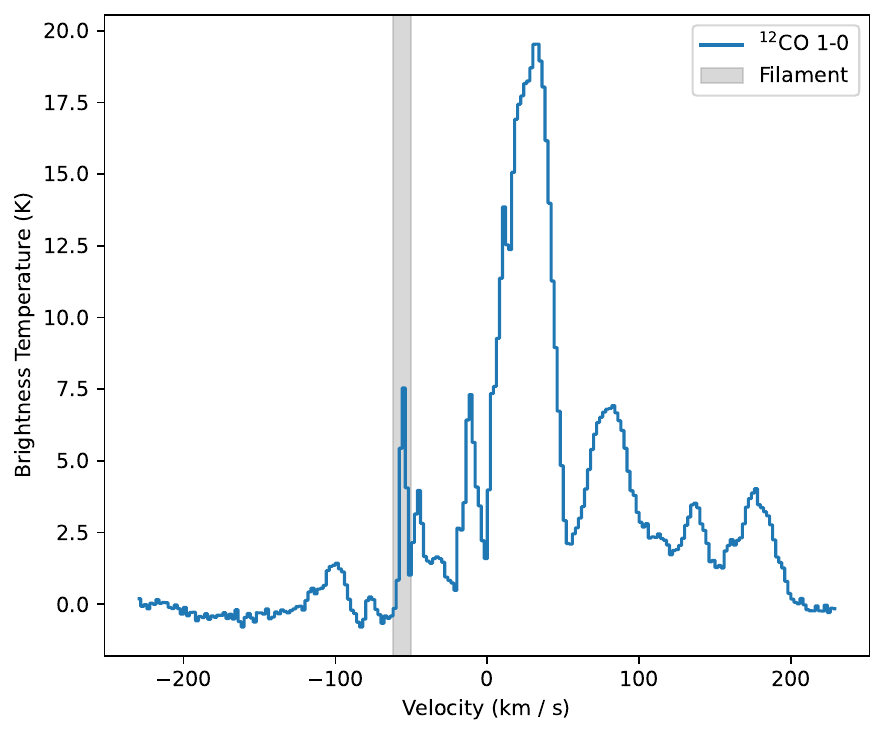}
    \caption{$\mathrm{^{12}CO}$ $J=1-0$ spectrum taken towards the filament taken over a $1.5~\arcmin$ aperture from the \citet{Tokuyama2019} survey of the Galactic Center. The region of the spectrum associated with the filament is highlighted in grey. Note that the filament is much narrower in line width than the spectral features at positive velocities.}
    \label{fig:filament-spectrum}
\end{figure}

\begin{table}
\caption{Filament Line Detections}
\centering
\begin{tabular}{ccc}
\hline
Line & Rest & ACES \\%& $\mathrm{A_{ul}}$ & Collisional Rates\footnote{From Leiden Atomic and Molecular Database \citep{lambda2005} accessed December 2024} & Critical \\
Name & Freq & 12m \\%&  & (T = {20}{K}) & Density\\
  & $\mathrm{GHz}$ & SPW \\%& $\mathrm{s^{-1}\, \times 10^{-6}}$ & $\mathrm{cm^{3}\, s^{-1}\, \times 10^{-11}}$ & $\mathrm{cm^{-3}\, \times 10^5}$ \\
 \hline
 \hline
CS $J=2\rightarrow1$ & 97.98095 & 33 \\%& 16.739 & 4.66 & 3.595 \\
HNCO $J=4\rightarrow3$ & 87.925238 & 31 \\%& 2.926 & 0.92 & 3.181 \\
$\mathrm{HNCO^+}$ $J=1\rightarrow0$ & 89.18852 & 29 \\%& 41.867 & 22.19 & 1.887 \\
SiO $J=2\rightarrow1$ & 86.84696 & 27 \\%& 29.374 & 19.41 & 1.513 \\
$\mathrm{H^{13}CO+}$ $J=1\rightarrow0$ & 86.7543 & 27 \\%& 7.152 & - & - \\
$\mathrm{HN^{13}C}$ $J=1\rightarrow0$ & 87.09085 & 27 \\%& 18.673 & - & - \\
SO $J_N=3_2\rightarrow2_1$ & 99.29987 & 33 \\%& 11.476 & 3.92\footnote{T={60}{K}} & 2.928 \\
$\mathrm{HC_3N}$ $J=11\rightarrow10$ & 100.0763 & 35 \\%& 77.7 & 5.18 & 15.0 \\
$\mathrm{H^{13}CN}$ $J=1\rightarrow0$ & 86.33992 & 25 \\%& 22.254 & - & - \\
\hline
\end{tabular}
\label{tbl:lines}
\end{table}

We determine the spectral properties of the filament, including its line of sight velocity and FWHM, using molecular line emission from the ACES data \citep{Longmore2026, Ginsburg2026, Walker2026, Lu2026, Hsieh2026}. 
At velocities centered at $-55~\kms$, we find several molecular lines showing evidence of bipolar molecular outflows as well as molecular lines faintly tracing the morphology of the filament's extinction itself.
In this part of the CMZ, clouds associated with the Galactic Center have positive velocities, while the only spiral arm feature in position-velocity space with a velocity of $-55~\kms$ toward the Galactic Center is the $\mathrm{3~kpc}$ arm \citep{dame2008}. 
The distance to the filament is further discussed in Section \ref{sec:distance}, but for these results we assume a distance of $\mathrm{5~kpc}$.

Velocity integrated intensity maps of the lines in Table \ref{tbl:lines} are shown in Appendix Figure \ref{fig:app-mom0}. The cloud is also detected in CO line emission in the \citet{Tokuyama2019} survey, as shown in Figure \ref{fig:filament-spectrum} where the velocity feature at the same line of sight velocity as the line emission detected in ACES is highlighted in the spectrum. 

% cloudc-jwst-2221/analysis/extract_spectra.py
We measure the velocity FWHM of the filament cloud by extracting an average spectrum from a $1.5~\arcmin$ radius circular aperture centered on the filament, then fitting a gaussian to the velocity component centered at $-55~\kms$. We measure a FWHM of $1.8~\kms$ in HNCO $J=4-3$, and using CO $J=1-0$ we measure a FWHM of $4.7~\kms$. 
Referencing the spectral feature in Figure \ref{fig:filament-spectrum}, we see that the highlighted feature associated with the filament has a smaller linewidth than any of the velocity features at positive velocities. These large linewidth features are associated with the CMZ, which has a positive velocity at the position of the filament. 

This implies that the filament is associated with a cloud in the disk of the Galaxy, most likely in the $\mathrm{3~kpc}$ arm, \referee{and not part of the CMZ.}

\subsection{Mass of the Filament} \label{sec:fil_mass}

\begin{figure}
    % lactea-filament/notebooks/extinction_map_extended.ipynb
    \centering
    \includegraphics[width=1\linewidth]{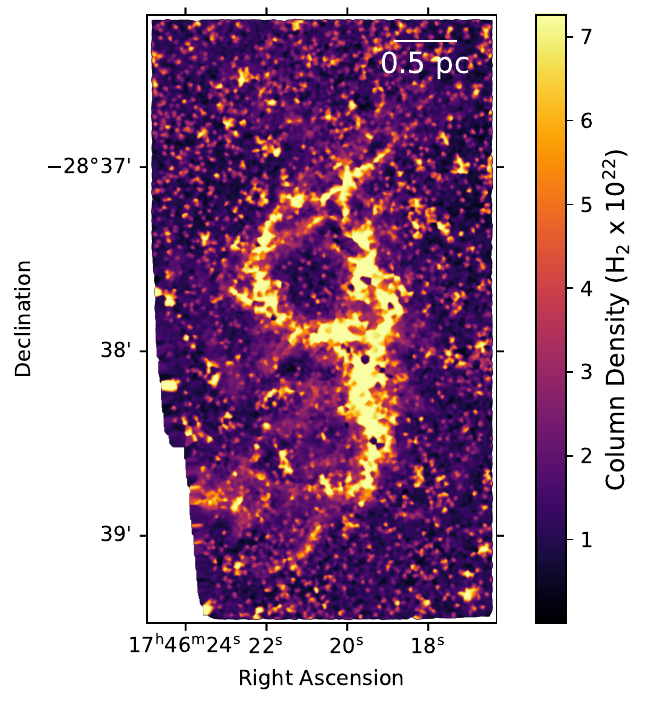}
    \caption{Column density map of the filament created using dust extinction measured with F182M and F410M.}
    \label{fig:column_density_map}
\end{figure}

%\noindent
\begin{table*} 
% notebooks/filament_mass_table.ipynb
\centering
\caption{Filament Mass Estimates}
\begin{tabular}{lcccccc}
\hline
Method\footnote{Using CO/$\mathrm{H_2}$=$2.5\times10^{-4}$} & $\mathrm{H_2}$ & $\mathrm{H_2}$ & $\mathrm{H_2}$ Column & $\mathrm{H_2}$ & CO Column & CO\\
 & Mass & M$_\mathrm{err}$\footnote{M$_\mathrm{err}$ is either the standard deviation (if reported as a single number) or corresponds to the minimum/maximum derived for different extinction curve adoptions as described in Section \ref{sec:fil_mass}.} & Density & N$_{err}$ & Density & N$_{err}$\\
 & $\mathrm{M_{\odot}}$ & $\mathrm{M_{\odot}}$ & $\mathrm{cm^{-2} \times 10^{22}}$ & $\mathrm{cm^{-2}\times 10^{22}}$ & $\mathrm{cm^{-2}\times 10^{17}}$ & $\mathrm{cm^{-2}\times 10^{17}}$ \\
 \hline
 \hline
$\mathrm{A}_\mathrm{V} \times 1.1 \times 10^{21}$ & 5600 & 3900 - & 2.0 & 1.7 & - & - \\
 & & 5600 & & & & \\
$^{12}$CO X-factor & 2400 & 430 & 0.80 & 0.15 & - & -\\%20 & 2.5\\
$^{13}$CO LTE & 340 & 62 & 0.11 & 0.02 & 2.9 & 0.53 \\
C$^{18}$O LTE & 1000 & 420 & 0.34 & 0.14 & 8.7 & 3.5 \\
CO Ice & 4700 & 4700 - & 1.7 & 1.3 & 40 & 33\\
 & & 8400 & & & & \\
PPMAP & 15000 & 260 & 2.5 & 0.04 & - & - \\
\hline
\hline
$\mathrm{A}_\mathrm{V} \times 1.1 \times 10^{21}$\footnote{The second half of this table is for only near the filament, cutting out the background.} & 3100 & 2200 - & 2.8 & 1.9 & - & - \\
 & & 3100 & & & & \\
$^{12}$CO X-factor & 940 & 160 & 0.83 & 0.14 & - & -\\%21 & 0.23 \\
$^{13}$CO LTE & 140 & 16 & 0.13 & 0.01 & 3.2 & 0.35 \\
C$^{18}$O LTE & 510 & 110 & 0.45 & 0.10 & 11 & 2.5 \\
CO Ice & 2600 & 2600 - & 2.3 & 1.4 & 58 & 34 \\
 & & 4400 & & & & \\
PPMAP & 5400 & 98 & 2.1 & 0.04 & - & - \\
\hline
\label{tbl:mass}
\end{tabular}
\end{table*}

% CO ice fraction 
% 

We measure the mass of the filament through several different strategies. The masses for each method are reported in Table \ref{tbl:mass}. 
The first half of the table has masses for the filament and the region around it (the full rectangular area covered e.g. by Figure \ref{fig:extinction_map}), while the second half of the table is limited to a closer-in region around filament, as shown in Figure \ref{fig:ccd}. 
For each mass measurement, we assume a molecular weight per particle of $\mathrm{H_2}$ of $\mu_{H_2}=2.8$ \citep{Kauffmann2008}. As the filament is associated with the $\mathrm{3~kpc}$ arm, we assume a distance of $\mathrm{5~kpc}$, which is discussed further in Section \ref{sec:distance}. We assume an abundance of $10^{-4}$ for CO relative to $\mathrm{H_2}$, but the effects of other CO abundances are shown in Figure \ref{fig:COice_Av}. 

\subsubsection{Extinction} \label{sec:ext_mass}
We use the relationship of $\mathrm{N_{H}}~(\mathrm{cm}^{-2}) = (2.21\pm0.09)~\times~10^{21} \mathrm{A_V}~(\mathrm{mag})$ between optical extinction $\mathrm{A_V}$ and hydrogen column density $\mathrm{N_{H}}$ from \citet{Guver2009} to estimate the mass of the filament. To get the relationship between $\mathrm{{N_{H_2}}}$ and extinction $\mathrm{A_V}$, we divide $\mathrm{N_{H}}$ in half. 
The process of measuring the extinction using [F182M]-[F410M] is detailed in Section \ref{sec:extinction}.
We multiply the extinction map in Figure \ref{fig:extinction_map} by the conversion factor, resulting in the column density map in Figure \ref{fig:column_density_map}. 
We measure the physical size of the pixels, and then sum over all of the pixels to measure the amount of $\mathrm{H_2}$ molecules. Then we multiply by the molecular weight per molecule to find the total mass of the filament with extinction. 
%We use the molecular weight per hydrogen molecule $\mu_{H_2}=2.8$ \citep{Kauffmann2008} to calculate the mass in each pixel
%We measure a mass of20326}{M_{\odot}} using the \texttt{CT06\_MWAvg} extinction law \citep{chiar2006}, as reported in Table \ref{tbl:mass}. 

We favor the \texttt{CT06} extinction law \citep{chiar2006} for the purposes of estimating the extinction and mass of the filament, as it better defines the dust extinction toward the Galactic Center. However, this extinction law is not necessarily the only or correct extinction law. To find the systematic error of the mass measurement due to different possible extinction laws, we find the mass of the filament using a variety of applicable Galactic extinction laws \citep{rieke1985, rieke1989, fritz2011, gordon2021, decleir2022} available through the \texttt{dust\_extinction} python package \citep{Gordon2024}. The mass reported in Table \ref{tbl:mass} is the mass measured using extinction $\mathrm{A_V}$ measured with \texttt{CT06}, which is also the highest mass from the list of extinction laws. The lowest mass was measured with \texttt{F11} \citep{fritz2011}, which includes a CO feature in the extinction curve. 

\subsubsection{X-factor}
We use the $\mathrm{^{12}CO}$ to $\mathrm{H_2}$ X-factor to find the mass of \referee{the} cloud using gas. 
With $\mathrm{^{12}CO}$ $J=1\rightarrow0$ observations from the %25 beam 
receiver (BEARS) on the Nobeyama Radio Observatory (NRO) \citep{Tokuyama2019}, we first isolated the $\mathrm{^{12}CO}$ emission from the filament by selecting only channels between the velocities of $-56~\kms$ to $-54~\kms$, as shown in Figure \ref{fig:filament-spectrum}. 
The physical resolution of these measurements is worse than JWST, with each pixel measuring $\mathrm{0.182~pc}$ ($\mathrm{7.50\arcsec}$) wide at the assumed distance to the filament of $\mathrm{5~kpc}$.
We then make a velocity integrated intensity map of the spectral cube data and cut out the same region used to estimate the mass with extinction. Using the \citet{strong88} X-factor of $\mathrm{2.3\times10^{20}~N(H_2)~cm^{-2} (K~\kms)^{-1}}$, we converted integrated intensity to column density. 
We then measured the mass by multiplying by the physical size of the pixels, summing them up, and applying the mean molecular weight factor. 
We measure the error by propagating the $\mathrm{0.3\times10^{20}~N(H_2)~cm^{-2} (K~\kms)^{-1}}$ error from the X-factor with the standard deviation of the integrated intensity map. 

The X-factor decreases as metallicity increases \citep{Bolatto2013}, which should make this mass measurement an upper limit. 
This is discussed further in Section \ref{sec:mass_recovery}.

\subsubsection{CO Isotopologue LTE}
We estimated the $\mathrm{H_2}$ column density and mass of the filament using CO isotopologue column densities by assuming that the gas in the cloud is optically thin for $\mathrm{^{13}CO}$ and $\mathrm{C^{18}O}$, and we assume the cloud is in local thermodynamic equilibrium. If the CO isotopologue lines are optically thick subthermally excited, then the real column density found through these methods is likely higher. The subthermal excitation case is unlikely in the case of the filament due to the high densities expected at the center of star forming clouds.

While we assume it is optically thin for these measurements, the $\mathrm{^{13}CO}$ $1\rightarrow0$ line is likely at least partially optically thick at the column densities observed in the filament, which would result in our mass measurement using $\mathrm{^{13}CO}$ being an underestimate.
The $\mathrm{C^{18}O}$ $1\rightarrow0$ line is unlikely to be optically thick, as the densities where this line is optically thick would result in all of the CO being frozen out of the gas phase. 
While our final measurement of mass from $\mathrm{C^{18}O}$ is multiple times higher than the mass measured from $\mathrm{^{13}CO}$ emission, we cannot necessarily use this as evidence that $\mathrm{^{13}CO}$ is optically thin, as it is degenerate with the assumed isotope ratios used later to convert to $\mathrm{^{12}C^{16}O}$.

We use Equation 79 from \citet{mangum15} to calculate the column density of the CO isotopologues 
\begin{equation}
    \begin{split}
    N^{thin}_{tot} = \left(\frac{3h}{8 \pi^3 S \mu^2 R_i}\right)
    \left(\frac{Q_{rot}}{g_u}\right) 
    \frac{\exp\left(\frac{E_u}{k_B T_{ex}}\right)}
    {\exp\left(\frac{h \nu}{k_B T_{ex}}\right)-1} \\
    \times
    \int \frac{T_R d\nu}{f \left(J_\nu(T_{ex}) - J_\nu(T_{bg})\right)}
    \end{split}
    \label{eq:mangum}
\end{equation}
 where $h$ is Planck's constant, 
$Q_{rot}$ is the rotational partition function, $g_u$ is the degeneracy, 
$E_u$ is the energy of the upper energy level, 
$k_B$ is the Boltzmann constant, 
$\nu$ is the frequency of the transition, 
the sum of relative intensities 
$R_i = 1$ for $\Delta J = 1$ transitions, 
$f$ is the filling factor assumed to be 1, 
$T_{ex}$ is the excitation temperature, 
$\int T_R d\nu$ is the integrated intensity, 
$T_{bg}$ is the temperature of cosmic microwave background,
$J_\nu(T)$ is the planck function,
the line strength $S=\frac{J_u}{2J_u+1}$ for linear molecules where $J_u$ is the upper energy level, 
and the value for the molecular electric dipole moment ($\mu$) is from the Jet Propulsion Laboratory (JPL) Molecular Spectroscopy database and spectral line catalog \citep{Pickett1998}. We also used constants from The Cologne Database for Molecular Spectroscopy \citep{Muller2001, Muller2005, Endres2016}. 
We assume a temperature of $10~\mathrm{K}$ for the filament, and use the integrated intensity of the filament in $\mathrm{^{13}CO}$ and $\mathrm{C^{18}O}$ $J=1\rightarrow0$ to calculate the column densities for both molecules. 
We expect a temperature of $10~\mathrm{K}$ because the filament is embedded in a molecular cloud and is not subject to strong radiation or turbulence. 
The peak intensity of the measured $\mathrm{^{12}CO}$ toward the filament is $10.8~\mathrm{K}$, so if we assume the $\mathrm{^{12}CO}$ is optically thick, then $T_B=T_{ex}$. 

We measure the mass of the filament using the CO isotopologues by converting the derived column densities of CO isotopologues into $\mathrm{H_2}$ column densities. 
To convert from CO isotopologues to $\mathrm{^{12}C^{16}O}$, we use abundance ratios for the isotopes of carbon and oxygen from \citet{Henkel1985, wilson94}, assuming the ratios $\mathrm{^{12}C/^{13}C}=53$ and $\mathrm{^{16}O/^{18}O}=327$ for the $\mathrm{4~kpc}$ molecular ring. 
We use an assumed CO abundance of $2.5\times10^{-4}$ to convert to $\mathrm{H_2}$ column density. 
The masses measured are reported in Table \ref{tbl:mass}. The column density measurements reported in the table are the average per pixel over the map, with the column density error being the standard deviation of the column densities across the map.

\subsubsection{CO Ice}
We estimated the $\mathrm{H_2}$ mass of the filament using CO ice. Figure \ref{fig:COice_map} shows the CO ice column density map of the filament. 

We convert this to mass by multiplying by an assumed CO abundance of CO/$\mathrm{H_2}$$=2.5 \times 10^{-4}$. Figure \ref{fig:COice_Av} shows why neither of the values of $10^{-4}$ \citep{Pineda2010} or $1.6\times10^{-4}$ \citep{Lacy2017} are appropriate CO abundances for the filament. Both underestimate the amount of CO relative to $\mathrm{H_2}$ present if our measurements of CO ice column density are correct. 

Finally, we multiply by the pixel size 
and sum the map. The mass measured with this method is shown in Table \ref{tbl:mass} and uses the \texttt{CT06} extinction law \citep{chiar2006}. Other extinction laws \citep{rieke1985, rieke1989, fritz2011, gordon2021, decleir2022} were used to find the range of masses included in the $M_{err}$ column, with \texttt{RRP89\_MWGC} being the most massive and \texttt{CT06} the lowest measured mass. We exclude the outlier measurement from the \texttt{F11} extinction law
due to it including the CO ice feature we are using. Using this extinction law results in an underestimate in the mass. 

\subsubsection{PPMAP}
We estimated the mass of the filament using \referee{the} PPMAP \referee{(Point Process MAPping) column density map of the Galactic Center} \citep{marsh2017}. PPMAP measures the \referee{dust} column density by fitting a dust spectral energy distribution (SED) using Hi-GAL, Herschel PACS and SPIRE. It assumes a dust opacity value of $\kappa_0 = $ $\mathrm{0.1~cm^2~g^{-1}}$ at $\mathrm{300~\micron}$, and $\beta = 2.0$. To measure the mass from the PPMAP data, we made a cutout of the column density map using the same region used for the filament previously. We summed over the pixels and multiplied by the physical area of the pixels at $\mathrm{5~kpc}$. Finally, we multiplied by the mean molecular weight $2.8$. \referee{We report the mass} in Table \ref{tbl:mass}, and we calculated the error using the PPMAP column density sigma map. 

There are some issues using PPMAP to estimate the mass of the filament. 
First, and most importantly, the resolution of PPMAP ($\sim12\arcsec$) is much worse than JWST, making them difficult to compare. 
Secondly, the filament is a feature of cold gas and dust interposed in front of diffuse, likely ionized gas in the Galactic Center. 
The PPMAP dust SED includes everything along the line of sight, so these background HII regions dominate the dust emission. 
While the filament is resolved in Herschel $70~\micron$, its small size means that it is difficult to identify in redder bands, and does not appear as an independent structure.
Additionally, while we kept the cutout region consistent with all other mass measurement techniques, the PPMAP cutout includes some pixels associated with Galactic Center dust ridge clouds, further increasing the mass measured. 
The PPMAP measurement should thus be considered an upper limit to the filament's mass. 

\subsection{Parent Cloud} \label{sec:parent}

\begin{figure}
    % lactea-filament/notebooks/filament_figs.ipynb
    \centering
    \includegraphics[width=\linewidth]{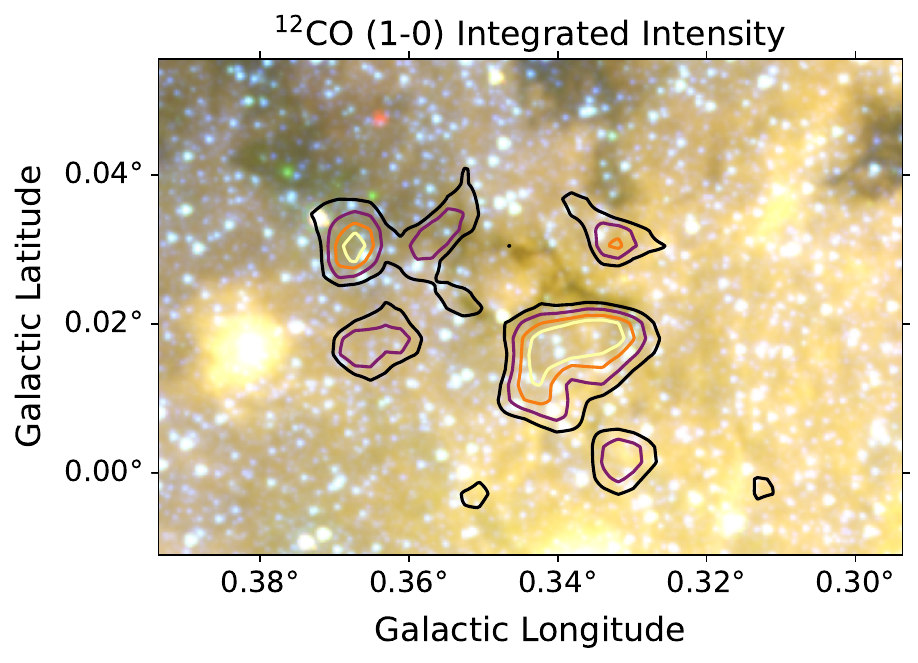}
    \includegraphics[width=\linewidth]{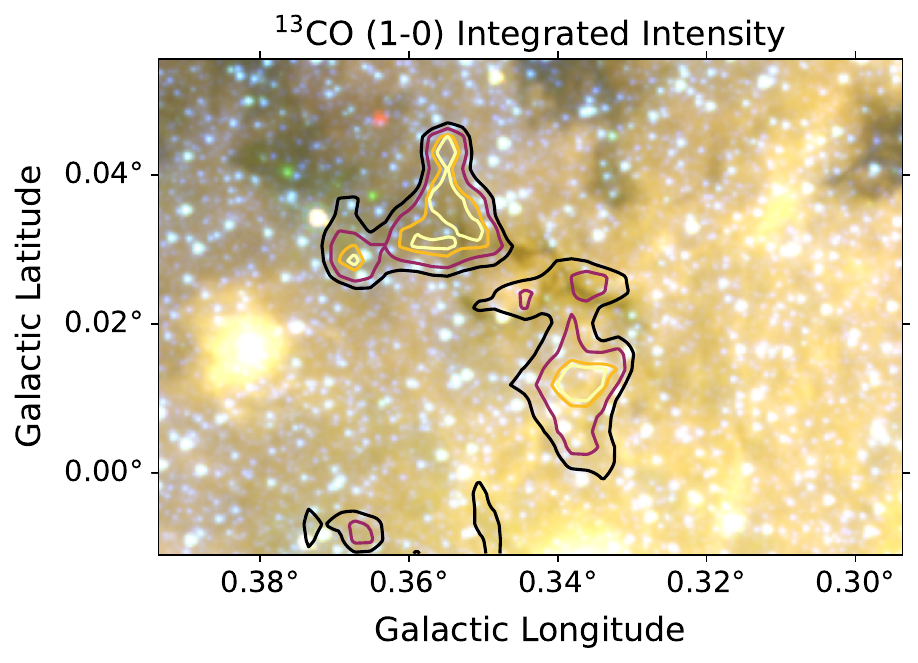}
    \includegraphics[width=\linewidth]{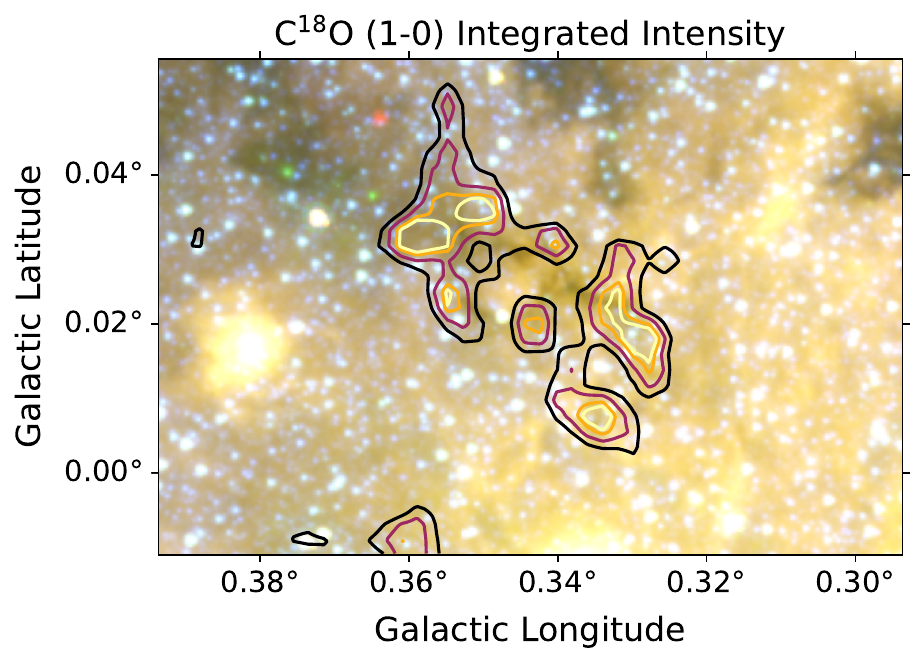}
    \caption{Color image is a Spitzer RGB image using I4, I3 and I1. The contours are NRO CO velocity integrated intensity maps of $\mathrm{^{12}CO}$, $\mathrm{^{13}CO}$, and $\mathrm{C^{18}O}$ taken from $-56~\kms$ to $-54~\kms$, with percentage levels of [90, 95, 98, 99] \citep{Tokuyama2019}.}
    \label{fig:co-mom0}
\end{figure}

The filament is located at the center of a larger gas cloud of $\mathrm{H_2}$ and CO. 

We estimate the mass of the parent cloud using the \citet{strong88} CO-to-$\mathrm{H_2}$ X-factor of $\mathrm{2.3\times10^{20}~cm^{-2}~(K~\kms)^{-1}}$ as an upper limit on the mass, assuming that the filament is typical of other Galactic disk clouds. 
We measure a mass of $\mathrm{7.4\times10^4~M_{\odot}}$ assuming a distance of $\mathrm{5~kpc}$.
As the filament is coincident with the Galactic Center on the sky, we also estimate the mass using an X-factor of $\mathrm{1.5\times10^{19}~cm^{-2}~(K~\kms)^{-1}}$ from \citet{gramze2023} for a lower mass estimate, assuming that the cloud is located in the Galactic Center $\mathrm{8~kpc}$ away or along the bar lanes. The mass measured using this lower X-factor is $\mathrm{1.2\times10^3~M_{\odot}}$.

The mass of the parent cloud measured with $\mathrm{^{12}CO}$ is still much lower than the mass measured using extinction, most likely due to CO freeze out in the cold interior of the cloud sending much of the gas phase CO into ice. 

As shown by Figure \ref{fig:co-mom0}, the parent CO cloud of the filament as observed by the Nobeyama single dish telescope \citep{Tokuyama2019} has distinctly different morphology and larger extent than the filament observed in the infrared. 

\subsection{Star Formation} \label{sec:sf}

\begin{figure*}
    % lactea-filament/notebooks/filament_outflows.ipynb
    \centering
    \includegraphics[width=\linewidth]{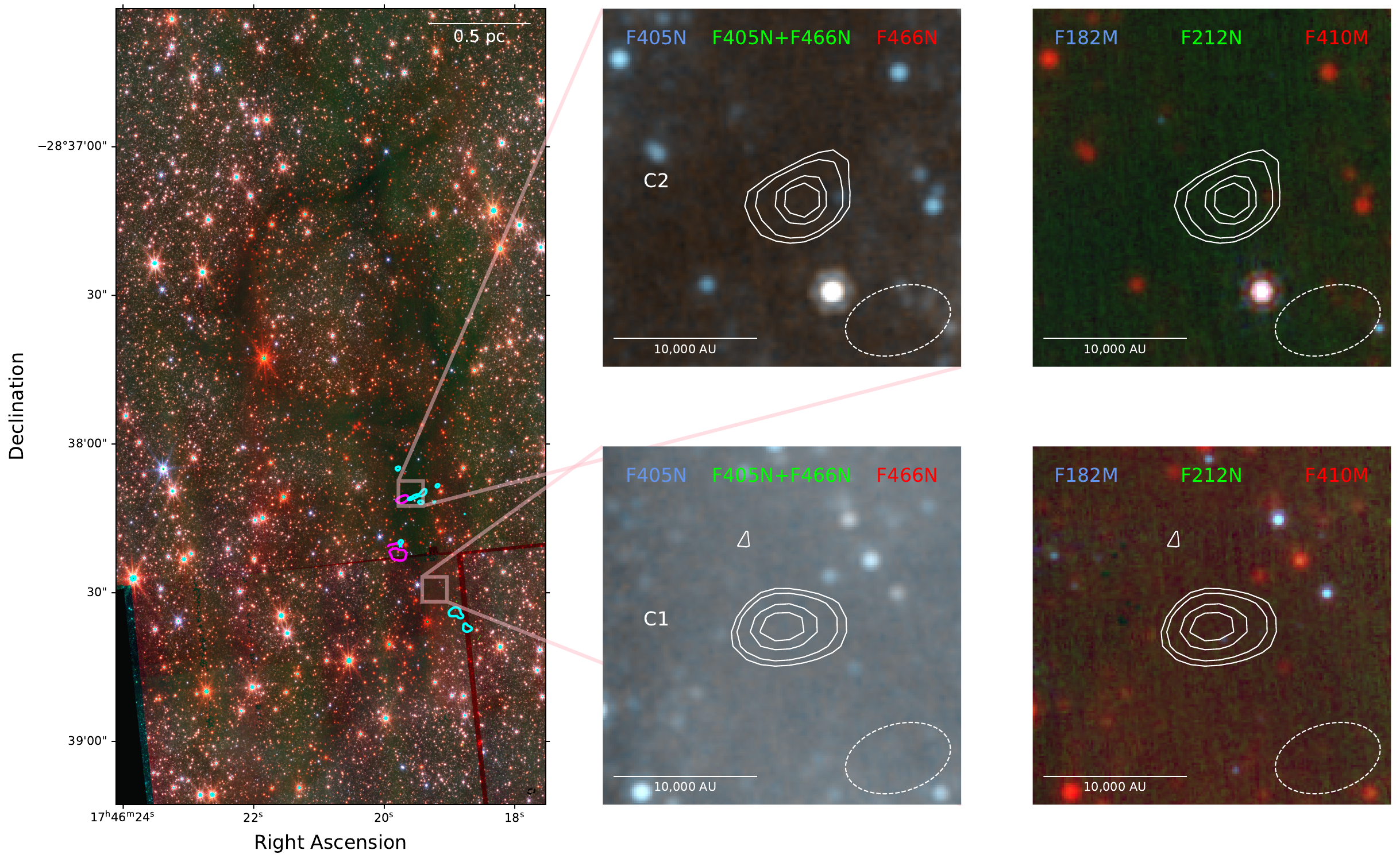}
    \caption{
    The left panel shows a three-color cutout of the filament using F410M, F212N, and F182M overlaid with ACES SiO $2\rightarrow1$ outflow contours. Note that there is no widespread emission in F212N, which corresponds to $\mathrm{H_2}$ and should be expected toward outflows. 
    The zoom-in panels show ACES Band 3 continuum contours on top of 3-color NIRCam cutouts. We do not observe any emission originating from these cores with JWST.}
    \label{fig:filament-zooms}
\end{figure*}

We find evidence of active star formation in the filament. In ACES Band 3, we identify two protostellar cores in continuum along with outflows in the line data centered at $-55~\kms$. The left panel of Figure \ref{fig:filament-zooms} shows the SiO $J=2-1$ outflows as contours overlaid on top of a three color image of the filament, with zoom-ins on the middle and right panels showing the contours of the Band 3 continuum emission for the two cores overlaid on three color images. We name the brighter of the two cores, the Southern core in Figure \ref{fig:filament-zooms}, core C1, while the dimmer core is C2. 

In the SCUBA-2 Galactic Center $850~\micron$ CO-corrected Compact Source Catalog \citep{Parsons2018}, a single submillimeter source G000.338+00.025 with an effective radius of $29\arcsec$ was detected at the location of the two cores identified in ACES. As both cores are included in the area of the effective radius, we conclude that the ACES ALMA data resolve G000.338+00.025 into two separate millimeter continuum sources. 

While the ACES observations of the cores do not resolve them, as these objects are associated with molecular outflows, they most likely have disks that are launching them. 
However, the objects do not have any cospatial JWST NIRCam emission. 
The cores are within the most heavily extincted region of the filament, and are thus deeply embedded. 
While the cores are associated with molecular outflows in the radio, there is no evidence of $2.12~\micron$ $\mathrm{H_2}$ emission in the F212N band, likely due to dust extinction. Observations with the F470N JWST filter would likely detect $\mathrm{H_2}$ emission associated with these outflows, as the longer wavelength filter pierces through more dust extinction. 

% SPICY YSOs
We also follow up YSO candidates from the SPICY catalog \citep{Kuhn2021}. We find 6 YSO candidates from the catalog in the region of the filament. Each of the candidates roughly aligns with at least one star, though some resolve into multiple. 
As the distance to these YSO candidates is not known, it is difficult to know if they are associated with the filament. 
The presence of the nuclear stellar disk in the background of the filament makes it difficult to verify if they are associated with that feature instead. 
None of these stars have notable features in JWST images or photometric colors, so we cannot confirm that these are YSOs. 

% Spitzer YSO candidates, aka the nebulous little guys
\begin{figure}
% YSOmass.ipynb
    \centering
    \includegraphics[width=\linewidth]{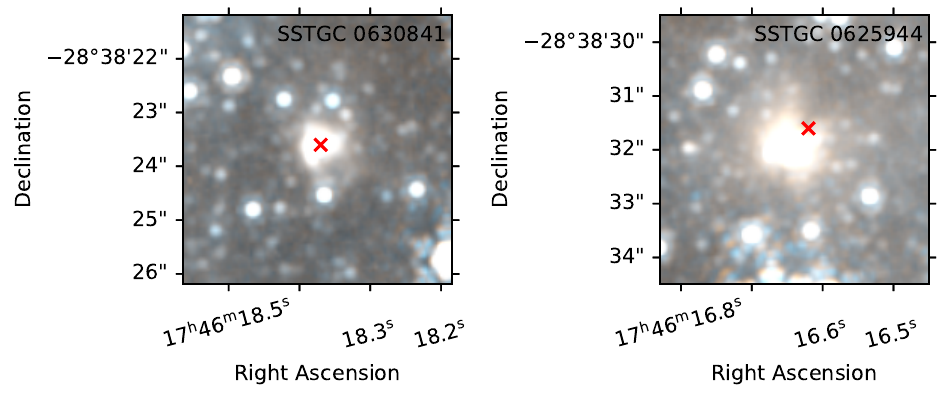}
    \caption{F405N and F466N image of two YSO candidates with surrounding nebulosity. The positions of the YSO candidate classifications are overplotted with red X's. Note that the two YSO candidates are red, brighter in F466N than F405N.}
    \label{fig:yso_candidates}
\end{figure}

We also find two YSO candidates from the Spitzer IRAC survey of the Galactic Center \citep{Ramirez2008}, SSTGC 0630841 and SSTGC 0625944, as shown in Figure \ref{fig:yso_candidates}. 
In JWST, the YSO candidates are surrounded by nebulous extended emission. They also appear red in F405N-F466N. The distance to these YSO candidates is difficult to determine, as they are not associated with molecular line emission in ACES bands, so we are unable to determine if these YSO candidates are associated with the filament. 
Spatially, they lie between the filament and dust ridge Cloud B. 
\citet{YusefZadeh2009} fits SSTGC 0630841 with an SED \citep{Robitaille2007} to measure its parameters, finding a mass of $7.7\pm1.0~\mathrm{M_\odot}$ and an extinction of $\mathrm{A_V}=33.4\mathrm{mag}$. 
A second YSO candidate near the filament is identified by \citet{YusefZadeh2009}, SSTGC 0630480. SSTGC 0630480 has no notable features in JWST images or photometric colors, so we cannot confirm that it is a YSO. 

\subsubsection{Continuum Mass}
We estimate the mass of the two protostellar cores detected in the filament in Band 3, which were presented in Section \ref{sec:sf}. 
As we currently only detect these sources in ALMA Band 3 emission, we forgo SED fitting and use grey body emission assuming optically thin dust to estimate the mass. 
Assuming a distance of $\mathrm{5~kpc}$ from Section \ref{sec:distance}, a temperature of $10~\mathrm{K}$, and a representative frequency of $\mathrm{92.45~GHz}$, we use the peak flux measured for the cores in the ACES continuum data to measure the mass, which results in an upper limit to the mass of the objects. We measure a flux of $\mathrm{0.7~mJy}$ for core C1 and a flux of $\mathrm{0.4~mJy}$ for core C2. We use a functional form of $\kappa$ that depends on frequency for the dust opacity $\kappa_\nu$:

\begin{equation}
    \frac{\kappa_\nu}{\mathrm{cm^{-1}~g}} = \kappa_0\left( \frac{\nu}{\nu_0} \right)^\beta
\end{equation}

Where $\beta$ is the power law index, $\nu$ is the frequency, 
$\kappa_0$ is the dust opacity per gram of $\mathrm{H_2}$ along the line of sight assuming a dust to gas ratio of 100, and
$\nu_0$ is the frequency where $\kappa_0$ was measured. We define $\beta=1.75$ and $\kappa_0=0.0114$ at $\nu_0=271.1~\mathrm{GHz}$ \citep{Ossenkopf1994}. We then calculate the mass $M$ with:

\begin{equation}
    \frac{M}{M_\odot} = \frac{S_\nu  d^2  c^2}{2 \kappa_\nu \nu^{2} k_B T}
\end{equation}

Where $S_\nu$ is the flux in mJy, $d$ is the assumed distance of $\mathrm{5~kpc}$, $c$ is the speed of light, $k_B$ is the Boltzmann constant, and $T$ is the temperature. 

We measure masses of $\mathrm{18.4~M_{\odot}}$ for core C1 and $\mathrm{10.5~M_{\odot}}$ for core C2 using a temperature of $10~\mathrm{K}$, which are upper limits to the masses of the cores. The masses of the cores decrease to $\mathrm{9.2~M_{\odot}}$ for C1 and $\mathrm{5.3~M_{\odot}}$ for C2 when assuming a temperature of $\mathrm{20~K}$. 

\subsubsection{Outflows}

\begin{figure*}
    % lactea-filament/notebooks/filaments_mom0_pv.ipynb
    \centering
    \includegraphics[width=0.95\linewidth]{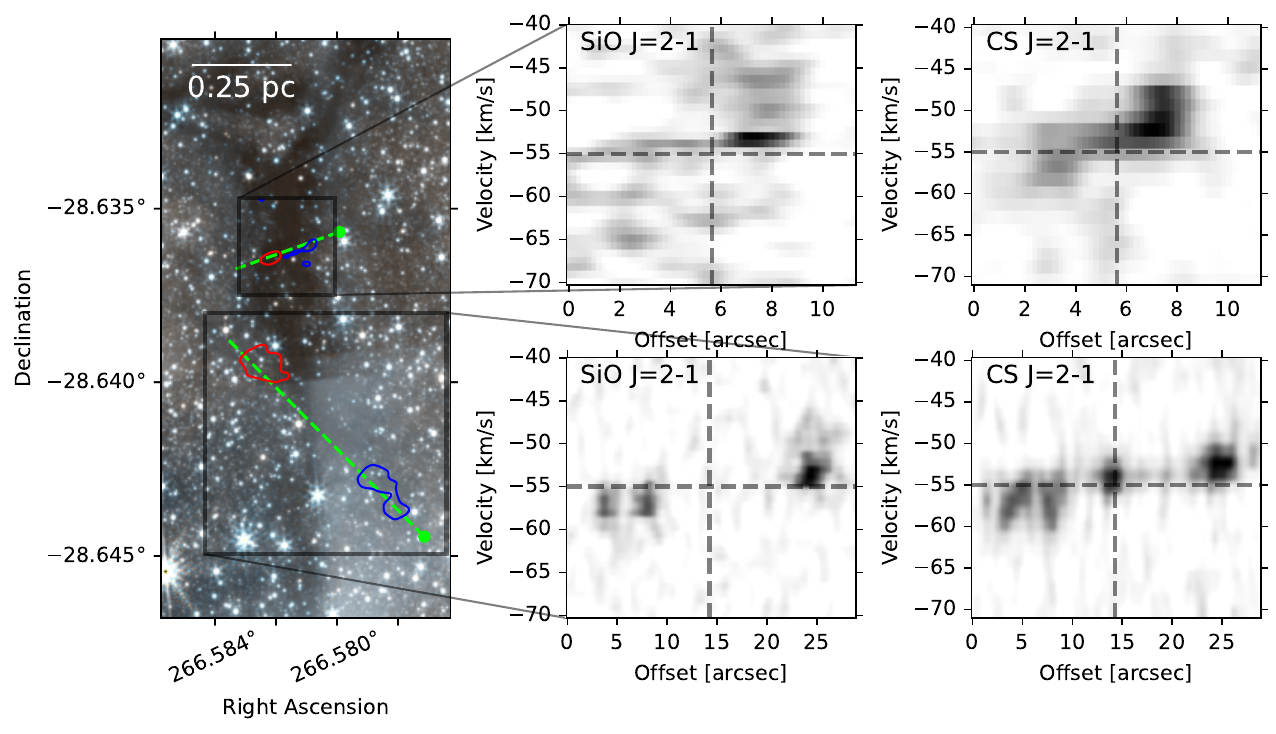}
    \caption{
    SiO $2\rightarrow1$ outflow contours are overlaid on top of a F466N and F405N RGB image of the filament zoomed in on the active star formation. The scale bar assumes a distance of $\mathrm{5~kpc}$.
    The bottom position-velocity diagrams are associated with core C1, and the top diagrams are for core C2. 
    The horizontal and vertical lines on the position-velocity diagrams indicate the positions of the two cores. 
    }
    \label{fig:filament-outflow-pv}
\end{figure*}

The left panel of Figure \ref{fig:filament-outflow-pv} shows the positions of the outflows relative to the filament and the cores from which the outflows emerge. The panels to the right of the figure show position-velocity diagrams taken across the outflows in SiO and CS. 
\referee{We detect many lines likely excited by the outflows, as shown in the spectra and integrated intensity maps in Appendix Figures \ref{fig:mol_spec} and \ref{fig:app-mom0}. 
Lines such as SiO $J=2-1$, which is only detected toward the outflows, represent shock tracers.
Other lines are often detected in multiple environments. 
Despite the many molecular lines identified in these outflows, we do not detect any extended emission in JWST F212N, which is centered on an $\mathrm{H_2}$ emission line excited by protostellar shocks such as these.
Figure \ref{fig:filament-zooms} includes F212N, but there is no $\mathrm{H_2}$ emission visible associated with the filament. The only extended green color associated with F212N is due to 1/f noise.}
The lack of $2.12~\micron$ emission is likely due to the outflows being behind heavy extinction. 
The outflows barely escape the most extincted part of the filament, likely reaching total extinctions (foreground+filament) of over $\mathrm{80~mag}$. 

\subsubsection{Other Feedback}

Notably, the filament seems to have a nearly circular hole at one end. While this is a very distinctive feature, currently our data do not suggest that this hole was formed due to feedback. There is no massive star, which would likely be very bright in the center of the hole, and the hole lacks any substantial increase in Br $\alpha$ emission in the F405N filter. There is no evidence of outflows present in this part of the filament, in ALMA or $\mathrm{H_2}$ emission in JWST F212N. 
No infrared YSOs are associated with the hole, most candidates seem to instead surround the IRDC. 
As for radio continuum sources, while there could be more lower mass stars being formed in the filament, it is unlikely that they would be able to blow out a hole in the filament this large without evidence in our current data. 
Most likely, this hole was created due to the filament branching into two arms, where projection effects make it seem as if it loops in on itself. However, we cannot rule out feedback as a formation mechanism. 

\section{Discussion} \label{sec:discussion}

%We have shown a filament that is backlit by thousands of Galactic Center stars, with a strong extinction and CO ice feature. 

The filament is backlit by thousands of stars. It features strongly in extinction, and shows signs of CO freeze-out. We find that different mass measurement techniques result in different masses for the filament. The filament shows signs of star formation, but the only confirmed YSOs are ALMA continuum sources with associated outflows. Now we discuss the distance estimate to the filament and how the fraction of CO locked in ice impacts mass estimates. 

\subsection{Distance Estimate} \label{sec:distance}
\referee{
Despite being in the same part of the sky as the CMZ, the filament is not associated with the Galactic Center. 
The number density of stars in the direction of the filament hints that it is closer and behind fewer stars than dust ridge clouds c and d. 
Its line of sight velocity is negative instead of positive like the CMZ clouds in the same region of the sky, where cloud C has two components of velocities $8~\kms$ and $39~\kms$ and cloud D has a velocity of $19~\kms$ \citep{Walker2025}.
We expect clouds in the CMZ to have broad lines because of the complex kinematic interactions and turbulent environment of the Galactic Center. Since this cloud has narrow lines, it is most likely not part of the CMZ. 
Its linewidth in CO is also lower than the CMZ clouds along the line of sight. 
%This implies that the filament is associated with a cloud in the disk of the Galaxy, most likely the $\mathrm{3~kpc}$ arm, and is not in the CMZ. 
}

\referee{
The filament's velocity of $-55~\kms$ matches the expected line of sight velocity for the $\mathrm{3~kpc}$ arm in the direction of the CMZ \citep{dame2008}. 
%% Move to velocity of the filament / distance estimate 
The spiral arm thought to be closest to the Galactic Center, the $\mathrm{3~kpc}$ arm, has an approximately elliptical ring shape \citep{Mulder1986, Fux1999, sanna2014}. 
Its near side was initially discovered using the $\mathrm{21~cm}$ HI line \citep{vanWoerden1957}, with the far side found using CO observations \citep{dame2008}. 
The nearside of the arm crosses $\ell = 0^{\circ}$ at a line-of-sight velocity of $-53~\kms$, with the far-side at $+56~\kms$ \citep{dame2008}. 
}

\referee{
The $\mathrm{3~kpc}$ arm has a uniquely large line-of-sight velocity compared to other spiral arms super-imposed on the Galactic Center. 
If the arm was orbiting perfectly along a circular orbit, its line of sight velocity toward the Galactic Center should be approximately $0~\kms$. Instead, its line of sight velocity toward the CMZ is $\sim$$-53~\kms$. 
%toward the Galactic Center is 
For this reason, kinematic distance measurements to the $\mathrm{3~kpc}$ arm falter. The orbits of material on the arm are unlikely to be circular, as the gravitational potential of the bar is dominant at smaller Galactic radii. 
Its velocity is thought to be due to the impact of bar dynamics causing non-circular motion in the gas close to the Galactic Center \citep{Mulder1986, Binney1991, Athanassoula1999, Li2022}. 
However, there remains some debate over whether this line-of sight velocity is instead a $\sim50~\kms$ expanding motion due to an ancient explosive event in the Galactic Center \citep{Oort1977, sofue2023}. 
%% End move to other sections 
}

While we know that the filament is not at the distance of the Galactic Center and is likely associated with the position-velocity feature of the $\mathrm{3~kpc}$ arm, there is a lack of distance measurements to the $\mathrm{3~kpc}$ arm in front of the Galactic Center, with one of the few being \citet{NoguerasLara2021a}, which measures a distance of $4.5\pm0.2~\mathrm{kpc}$. 
Indeed, distance measurements directly toward the Galactic Center are generally lacking. Since the shape of the $\mathrm{3~kpc}$ arm is thought to be approximately elliptical due to the influence of the Galactic bar, the distance to the $\mathrm{3~kpc}$ arm is most likely not exactly $\mathrm{5~kpc}$.

We attempt to estimate the distance to the filament, as explained in Appendix Section \ref{app:dist}. 
We first used a stellar density measurement of stars in front of the filament to compare to a TRILEGAL stellar population model, but we found that the stellar density of stars observed with JWST exceeds that of the model. 

We also used a couple of optical surveys to put a lower limit on the distance. Querying the Gaia survey \citep{Gaia2016, Gaia2023} for stars with distances \citep{Lindegren2021, Gaia2021} within our field, we found matches to only stars within $\mathrm{1~kpc}$. There were too few stars in our field matched with Gaia to make a distance estimate, as there were no stars behind or within the filament in this catalog. 

We also used the DECaPS2 (Decaps) survey \citep{Schlafly2018, Saydjari2023}, which is deeper and includes more bands than Gaia. However, after matching the stars with distances to our catalog, we found that the distances were not well matched to the extinction dominated color observed with JWST. Additionally, the matched stars had extinctions lower than those observed in or behind the filament. We find that the most likely explanation is resolution differences. One star in Decaps can be many stars in JWST. 

The failure of these methods of measuring the distance to the filament highlights a need for a method of measuring the distances toward objects in front of the CMZ. In other parts of the Galaxy, the proper motion of maser emission has been used to measure the distance to star formation regions \citep{sanna2014, reid2019} and associate them with well studied spiral arms. The filament has a line of sight velocity association with the $\mathrm{3~kpc}$ arm, has two known young stellar objects with molecular outflows, and has no detections in previous $\mathrm{H_2O}$ maser surveys \citep{Ladeyschikov2022L, Ward2024}
%\todo{CITE DYLAN WARD'S PAPER IF IT IS SUBMITTED B4 MINE}. 
Therefore, we assume a distance of $\mathrm{5~kpc}$, though this distance estimate is most likely not accurate.

\subsection{Abundance Implications}
As shown by Figure \ref{fig:COice_Av}, if we assume that the abundance of CO/$\mathrm{H_2}$$=10^{-4}$ \citep{Pineda2010} and $\mathrm{N}_\mathrm{H} = 1.1 \times 10^{21} ~\mathrm{A}_{\mathrm{V}}~\mathrm{(mag)}$  \citep{Guver2009}, a substantial number of stars detected behind the filament have CO ice column densities that surpass the case where 100\% of the CO is in ice. 
The most likely reason for this discrepancy, as shown by the additional red lines, is that the abundance of CO is higher toward the filament. 

Indeed, while the \citet{smith_cospatial_2025} data over-plotted in Figure \ref{fig:COice_Av} have been shifted in $\mathrm{A}_{\mathrm{V}}$, the points follow a trend that levels off at a CO ice column density matching CO/$\mathrm{H_2}$$=1.6 \times 10^{-4}$ reported by \citet{Lacy2017}. The \citet{smith_cospatial_2025} points level off at a lower column density than the filament, pointing toward the clouds having different CO abundances. 

We use CO/$\mathrm{H_2}$$=2.5\times10^{-4}$ to measure the masses estimated using CO isotopologues and CO ice in Table \ref{tbl:mass}. 
A lower abundance, such as CO/$\mathrm{H_2}$$=1.6\times10^{-4}$, brings the masses derived from the CO isotopologues more into line with the mass measured with the CO X-factor, but a higher abundance, such as CO/$\mathrm{H_2}$$=5\times10^{-4}$, is what best fits the measured CO ice column density shown in Figure \ref{fig:COice_Av}. 
If the CO abundance is as high as it needs to be to fit the CO ice column density measurements, then there is a substantial amount of CO gas measured via isotopologues missing. While the CO isotopologue lines could be underestimating the amount of mass present due to the lines being more optically thick than we assume, they are also likely affected by CO freeze-out, which is explored in the next section. 

\subsection{CO Ice Fraction} \label{sec:ice_frac}

We measure the amount of CO present in the cloud with two methods, measuring the CO ice column density and measuring the CO gas column density using LTE and isotope ratio assumptions. 
These are direct measurements of how much CO is in gas versus ice along the line of sight. 

Table \ref{tbl:mass} shows that the column density of CO in ice is $\sim5-20\times$ the amount of CO gas. This shows that substantial freeze-out of CO occurs in the filament, where only about $7-22\%$ of the CO is present in the gas phase. This measurement heavily relies on the molecular abundances and all of the other systematic uncertainties for the CO ice measurement mentioned in Section \ref{sec:ice_uncertain}. Nonetheless, this measurement implies that the majority of CO is locked in ice in the filament. 

Since CO ice does not contribute to the emission produced by gas phase CO, the $\mathrm{H_2}$ mass we measure using gaseous CO is smaller than the amount of mass present for infrared dark clouds. 
As the gas that forms stars is cold and dense, the perfect environment for ice formation, the impact of the CO ice fraction will be mostly felt in star forming clouds where ice is present.

\subsection{$\mathrm{H_2}$ Mass Fraction} \label{sec:mass_frac}

\begin{table}[]
    \centering
    \caption{$\mathrm{H_2}$ Mass Ratios}
    \begin{tabular}{ccc}
    \hline
        Method & Mass Ratio & Err \\
    \hline
    \hline
       CO Ice / (CO Ice  & 0.66\footnote{CO/$\mathrm{H_2}$$=2.5\times10^{-4}$} & 0.50\footnote{CO/$\mathrm{H_2}$$=5\times10^{-4}$} - 0.83\footnote{CO/$\mathrm{H_2}$$=10^{-4}$} \\
        \hspace{1mm}+CO X-factor) & & \\
       CO X-factor / & 0.42 & 0.42 - 0.60 \\
        $\mathrm{H_2}$ from $\mathrm{A_V}$ & & \\
        (CO Ice +CO X-factor) / & 1.26$^\mathrm{a}$ & 0.84$^\mathrm{b}$ - 2.51$^\mathrm{c}$ \\
        $\mathrm{H_2}$ from $\mathrm{A_V}$ & & \\
        (CO Ice +CO X-factor) / & 0.46$^\mathrm{a}$ & 0.31$^\mathrm{b}$ - 0.92$^\mathrm{c}$ \\
        PPMAP  & & \\
    \hline
       CO Ice / (CO Ice\footnote{The second half of this table is for only near the filament, cutting out the background.}  & 0.74$^\mathrm{a}$ & 0.58$^\mathrm{b}$ - 0.88$^\mathrm{c}$ \\
        \hspace{1mm}+CO X-factor) & & \\
       CO X-factor / & 0.30 & 0.3 - 0.42 \\
        $\mathrm{H_2}$ from $\mathrm{A_V}$ & & \\
        (CO Ice +CO X-factor) / & 1.32$^\mathrm{a}$ & 0.72$^\mathrm{b}$ - 2.38$^\mathrm{c}$ \\
        $\mathrm{H_2}$ from $\mathrm{A_V}$ & & \\
        (CO Ice +CO X-factor) / & 0.66$^\mathrm{a}$ & 0.4$^\mathrm{b}$ - 1.38$^\mathrm{c}$ \\
        PPMAP  & & \\
    \hline
    \end{tabular}
    \label{tab:mass_ratio}
\end{table}

The amount of $\mathrm{H_2}$ mass in the filament measured with dust extinction is larger than the mass measured using the X-factor or CO isotopologues. 

The X-factor decreases with increasing metallicity \citep{Bolatto2013, Gong2020, Kohno2024, Bisbas2025}. The metallicity in the Galactic Center is higher than the solar neighborhood \citep{Nandakumar2018}, so it decreases radially. The filament is in front of the CMZ, but it is not in the solar neighborhood, so it should have a higher metallicity and thus a smaller X-factor. Thus, the mass measured with the local X-factor should be an upper limit. However, the $\mathrm{H_2}$ mass measured with the X-factor, as shown in Table \ref{tbl:mass}, is substantially lower than the mass measured using dust extinction. 

Both of the $\mathrm{H_2}$ masses estimated using $\mathrm{^{13}CO}$ and $\mathrm{C^{18}O}$ LTE assumptions are also much lower than expected from the dust extinction. While this could be due to the lines being optically thick when we assume they are optically thin, the lack of mass estimated using the $\mathrm{^{12}CO}$ X-factor points to another explanation. 

The most likely cause of the discrepancy between the masses measured using CO emission and dust extinction is substantial CO freeze-out. Cold, dense clouds where stars form have the conditions where CO falls out of the gas phase and freezes onto dust grains. 

Table \ref{tab:mass_ratio} shows different $\mathrm{H_2}$ mass ratios for methods of estimating the mass of the filament. The first row shows that over 50\% of the mass as traced by CO is from CO ice, which is an underestimate of the amount of freeze-out due to the X-factor behavior mentioned above. 

\subsection{Mass Recovery} \label{sec:mass_recovery}
The mass measured using extinction is several times larger than the mass measured using the X-factor. Table \ref{tab:mass_ratio} shows that only $\sim$42-60\% of the mass measured with $\mathrm{A_V}$ is recovered using CO gas estimates for the total mass. 
This implies that the masses of resolved cold, dense, star-forming clouds measured with the X-factor are underestimated, as CO freeze-out reduces the amount of CO in the gas phase. 
This freeze-out of star-forming gas \reftwo{could be} one of the contributing factors to the observed scatter in the Kennicutt-Schmidt relation. 

As discussed in Section \ref{sec:ice_frac}, most of the CO in the filament is locked away in the ice phase, leading to the discrepancy between the measured masses. When we add the mass measured using the amount of CO locked up in ice, we estimate the mass measured with extinction $\mathrm{A_V}$, with the total mass measured with CO gas and ice being 152-161\% of the mass measured with extinction $\mathrm{A_V}$.
Table \ref{tab:mass_ratio} shows the ratio of the total mass measured with CO gas and ice to the mass measured with extinction $\mathrm{A_V}$. \reftwo{While CO freeze-out is affecting CO rotational line emission, the degree of uncertainty in mass recovery is less than a factor of two for the filament.}

In Section \ref{sec:parent}, we measured a total mass of \reftwo{the parent cloud of} $\mathrm{7.4\times10^3~M_\odot}$ using the X-factor, with an average surface density of $150 \pm40~\mathrm{M_\odot pc^{-2}}$.
\reftwo{If the metallicity of the filament is truly higher than solar, then we expect that the X-factor is lower, and the mass of the filament cloud as traced by CO gas is even lower than expected from dust extinction measurements.}
When we add the mass inferred from CO locked in ice to the mass measurement using the standard X-factor, the total mass of the parent cloud becomes $\mathrm{1.3\times10^4~M_\odot}$.

\section{Conclusion} \label{sec:conclusion}

\referee{
We observe an IRDC filament with signs of active star formation backlit by the Galactic Center with JWST NIRCam. 
We map the filament in extinction $\mathrm{A_V}$ and CO ice column density.
While we cannot accurately measure the distance to the filament, we can place it foreground of the Galactic Center, confirming that the near side of the $\mathrm{3~kpc}$ arm passes in front of the Galactic Center. 
Column density measurements of CO ice in the filament are higher than expected for local measurements of CO abundance, implying that the CO abundance for the filament is higher and that there is a Galactic gradient of CO abundance dependent on metallicity.}

\referee{
We estimate the mass of the filament. 
Mass estimates with various different techniques imply that 50-88\% of the CO in the filament is frozen out of the gas phase and is in CO ice. 
However, the mass estimate from combining mass measured with both CO gas and ice \referee{overestimates} of the mass measured from the dust using extinction $\mathrm{A_V}$.
Measuring the total mass of the parent cloud by including both the mass derived from the X-factor and the CO ice column density approximately doubles the mass. 
}

\referee{
For dense gas in the Milky Way and other galaxies, CO also likely freezes out, throwing off mass measurements using CO gas emission. We find evidence that the CO abundance in the filament is greater than for local clouds, implying that enhanced gas metallicity in the inner Galaxy drives increased CO abundance.
} 

\begin{acknowledgments}
\referee{We thank the referee their constructive comments that have improved this manuscript.} 
Optical constants data for CO ice were retrieved from the Optical Constants database. MCS acknowledges financial support from the European Research Council under the ERC Starting Grant ``GalFlow'' (grant 101116226) and from Fondazione Cariplo under the grant ERC attrattivit\`{a} n. 2023-3014.
This work is based in part on observations made with the NASA/ESA/CSA James Webb Space Telescope. The data were obtained from the Mikulski Archive for Space Telescopes at the Space Telescope Science Institute, which is operated by the Association of Universities for Research in Astronomy, Inc., under NASA contract NAS 5-03127 for JWST. These observations are associated with program \#2221.
Support for program \#2221 was provided by NASA through a grant from the Space Telescope Science Institute, which is operated by the Association of Universities for Research in Astronomy, Inc., under NASA contract NAS 5-03127.
AG acknowledges support from the NSF under grants AAG 2206511 and CAREER 2142300.
NB acknowledges support from the Space Telescope Science Institute via grant No. JWST-GO-05365.001-A. 
E.A.C.\ Mills  gratefully  acknowledges  funding  from the National  Science  Foundation  under  Award  Nos. 1813765, 2115428, 2206509, and CAREER 2339670.
F.N.-L. gratefully acknowledges financial support from grant PID2024-162148NA-I00, funded by MCIN/AEI/10.13039/501100011033 and the European Regional Development Fund (ERDF) “A way of making Europe”, from the Ramón y Cajal programme (RYC2023-044924-I) funded by MCIN/AEI/10.13039/501100011033 and FSE+, and from the Severo Ochoa grant CEX2021-001131-S, funded by MCIN/AEI/10.13039/501100011033.
BALG is supported by the German Research Foundation (DFG) in the form of an Emmy Noether Research Group - DFG project \#542802847 (GA 3170/3-1)."
C.\ Battersby gratefully acknowledges funding from National Science Foundation under Award Nos. 2108938, 2206510, 2414862, and CAREER 2145689, as well as from the National Aeronautics and Space Administration through the Astrophysics Data Analysis Program under Award ``3-D MC: Mapping Circumnuclear Molecular Clouds from X-ray to Radio,” Grant No. 80NSSC22K1125 as well as participation in the PRIMA project under Grant No. 80NSSC25K7944.

\end{acknowledgments}

\vspace{5mm}
\textbf{Facilities:} JWST, ALMA

\software{
This research has made use of the following software projects:
    \href{https://github.com/spacetelescope/jwst?tab=readme-ov-file}{JWST Pipeline} \citep{bushouse_2025_17101851},
    \href{https://github.com/spacetelescope/webbpsf}{WebbPSF} \citep{Perrin2014}, 
    \href{https://astropy.org/}{Astropy} \citep{astropy2022},
    \href{https://matplotlib.org/}{Matplotlib} \citep{Hunter2007},
    \href{http://www.numpy.org/}{NumPy} \citep{harris2020array},
    \href{https://github.com/astropy/regions?tab=readme-ov-file}{regions} \citep{larry_bradley_2024_13852178},
    \href{https://scipy.org/}{SciPy} \citep{2020SciPy},
    \href{https://casa.nrao.edu/}{CASA} \citep{CASA2022},
    \href{https://github.com/radio-astro-tools/spectral-cube}{spectral-cube} \citep{adam_ginsburg_2019_3558614},
    \href{https://github.com/radio-astro-tools/radio-beam}{radio-beam} \citep{eric_koch_2025_15677957},
    \href{https://github.com/keflavich/dust_emissivity}{dust\_emissivity},
    and
    the NASA's Astrophysics Data System.
}

\appendix

\section{Labeled Image}

\begin{figure*}
    \centering
    \includegraphics[width=\linewidth]{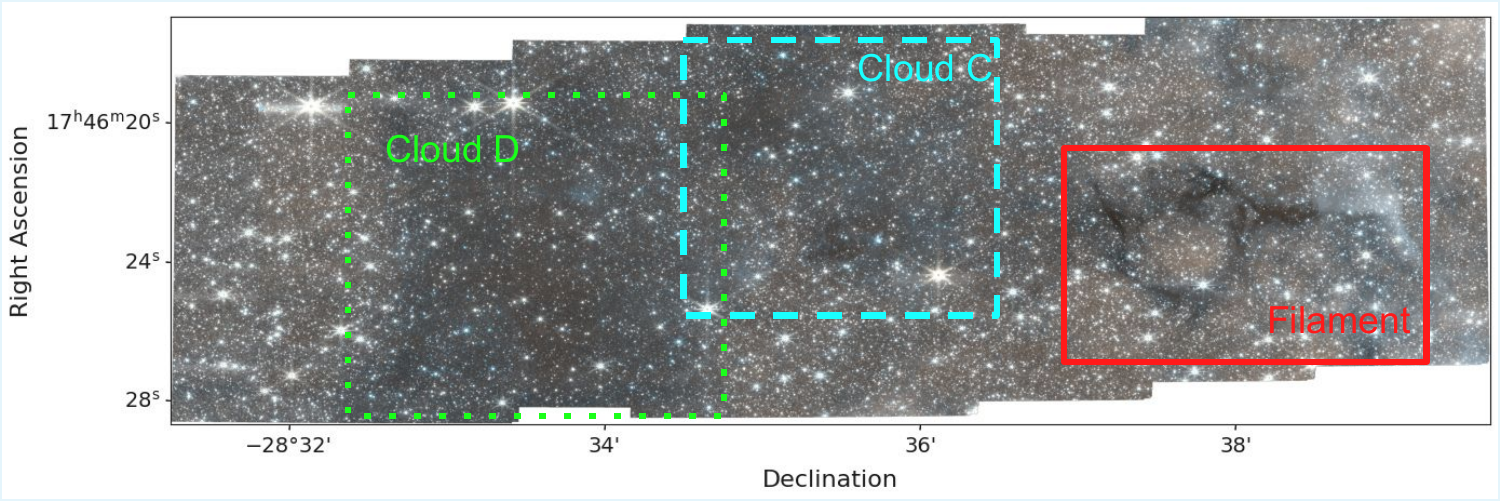}
    \caption{\referee{Figure \ref{fig:jwst-fov} with the most distinct infrared dark clouds labeled.}}
    \label{fig:cropped}
\end{figure*}

\referee{
Figure \ref{fig:cropped} shows a three color image of the region observed with JWST. This figure is the same as Figure \ref{fig:jwst-fov}, but with dust ridge clouds C and D and the filament labeled.
}
\section{Filtered Images}

\begin{figure*}
    % notebooks/MedianFilterBackground_CloudC.ipynb
    % notebooks/perfilt.ipynb
    \includegraphics[width=\textwidth]{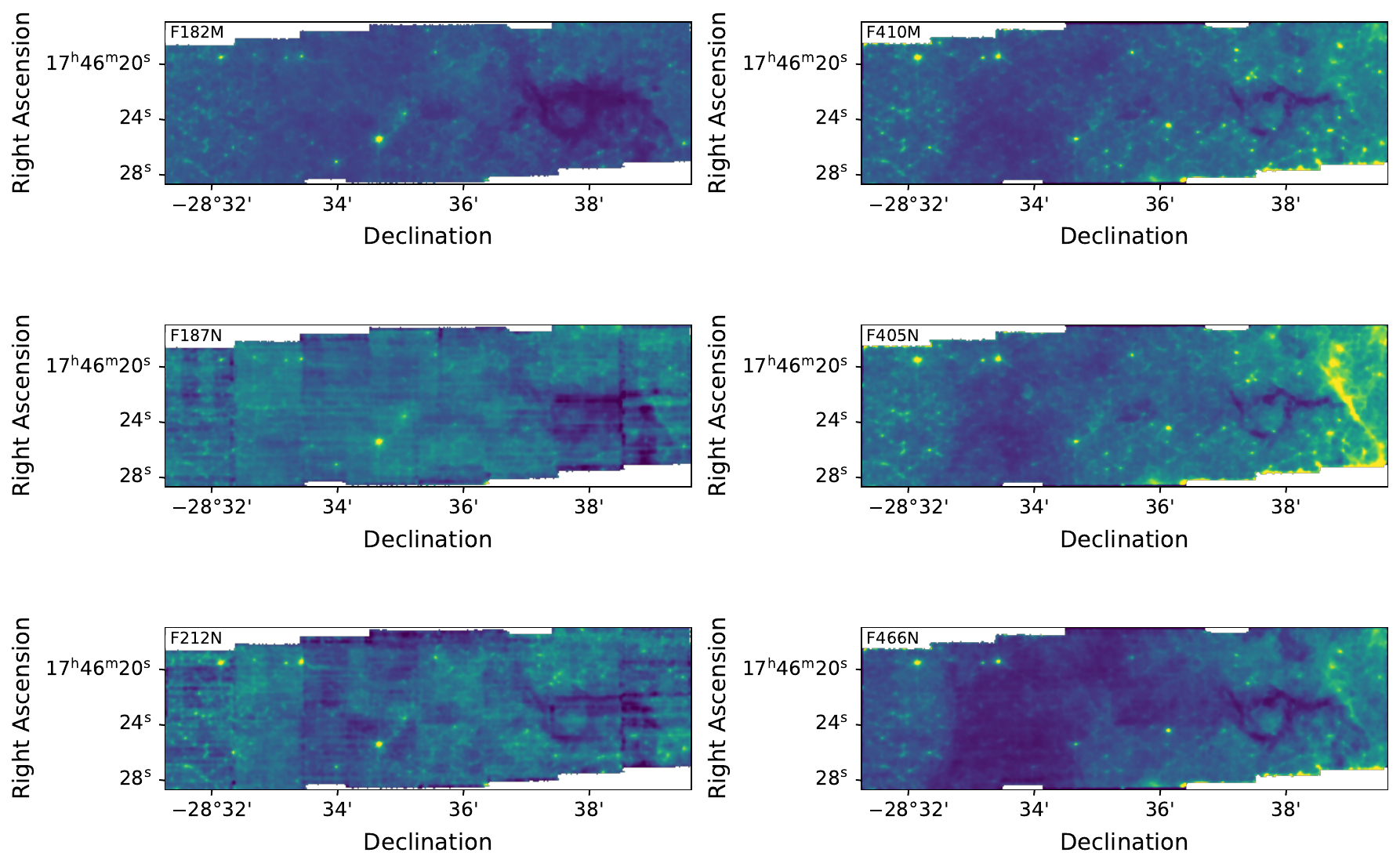}
    \caption{JWST NIRCam images before 1/f noise removal smoothed over with a circular 10th percentile filter to make images that are used to add back large scale structure.}
    \label{fig:app-perfilt}
\end{figure*}

Figure \ref{fig:app-perfilt} shows the images created by applying a circular percentile filter to images made with no 1/f noise removal techniques applied. The images show the large scale structure of the field, especially the differences in the extent of the filament in ths short and long wavelength filters. Dust ridge clouds c and d are also distinguishable in the long wavelength filters, and the F405N image has a prominent diffuse HII region. The F187N and F212N images have some residual 1/f noise that survived the filtering process. 

\section{\referee{Molecular Spectra}}

\begin{figure*}
    \centering
    %/orange/adamginsburg/jwst/cloudc/lactea-filament/lactea-filament/notebooks/filaments_mom0_pv.ipynb
    \includegraphics[width=0.95\linewidth]{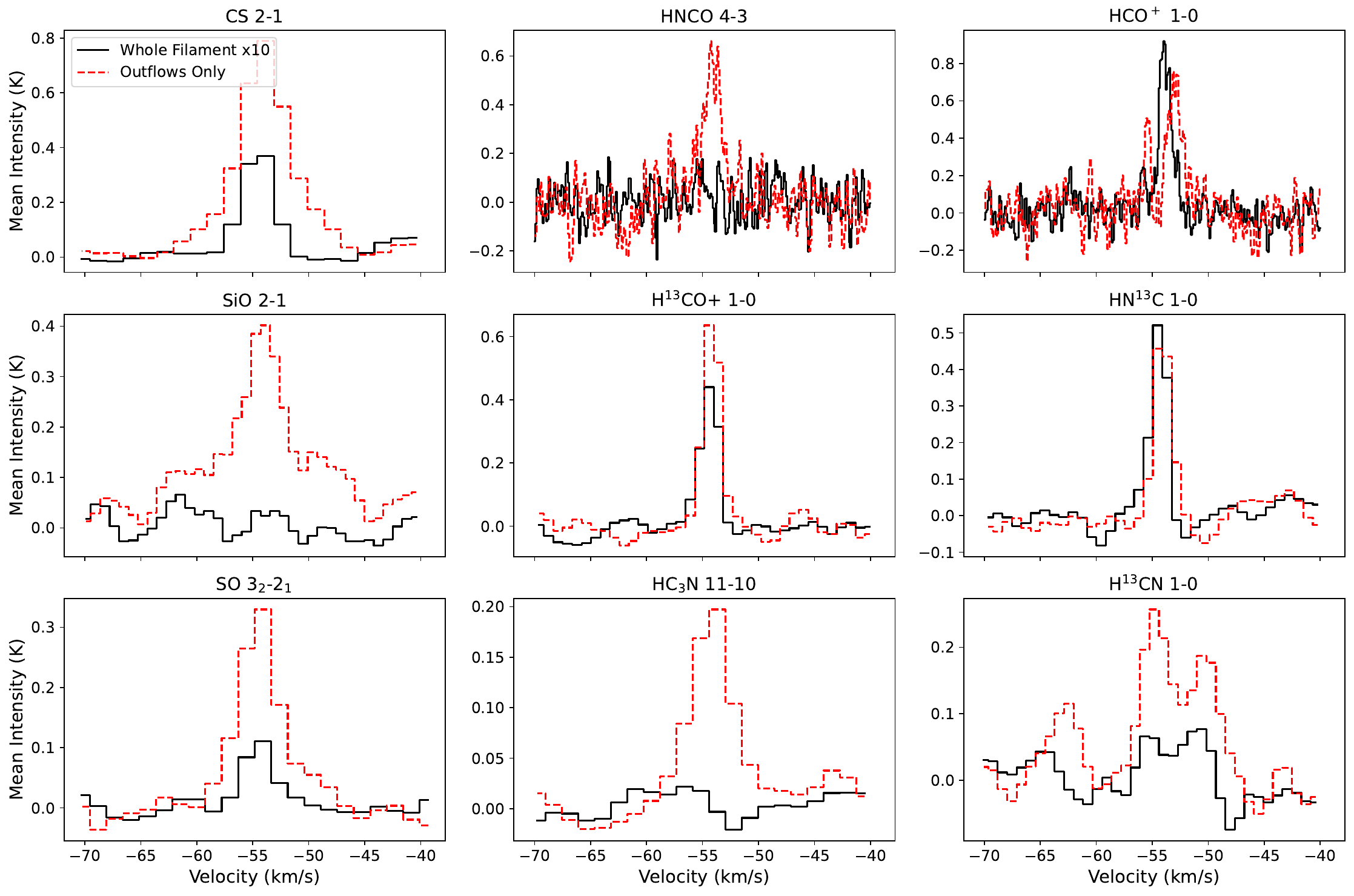}
    \caption{\referee{Average spectra from lines detected toward the filament as listed in Table \ref{tbl:lines}. The black lines are the spectra over the entire region of the filament, multiplied 10x, to be comparable with the red lines representing the spectra only toward the outflows.}}
    \label{fig:mol_spec}
\end{figure*}

\referee{Figure \ref{fig:mol_spec} shows the spectra of the lines detected toward the filament. The black lines represent average spectra taken over the entire region of the filament, while the red lines are average spectra taken only toward the outflows. As the black spectra are averaged over a larger area, we multiply them by 10x to more readily compare them with the intensity of the outflow spectra.}

\section{Moment Maps}

\begin{figure*}
    % filaments_mom0_pv.ipynb
    \includegraphics[width=0.33\textwidth]{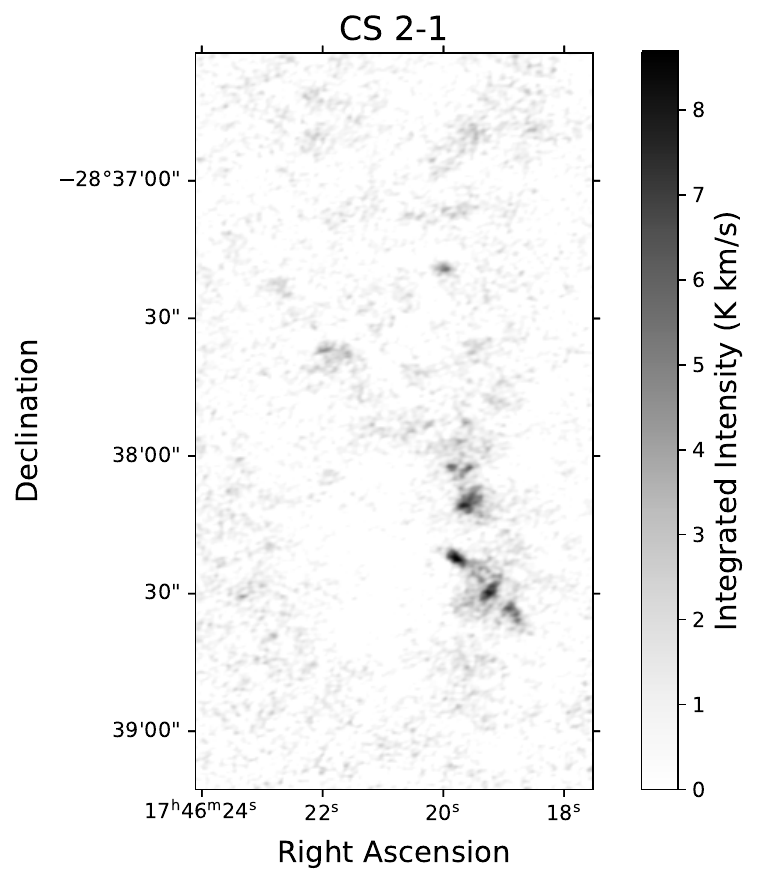}
    \includegraphics[width=0.33\textwidth]{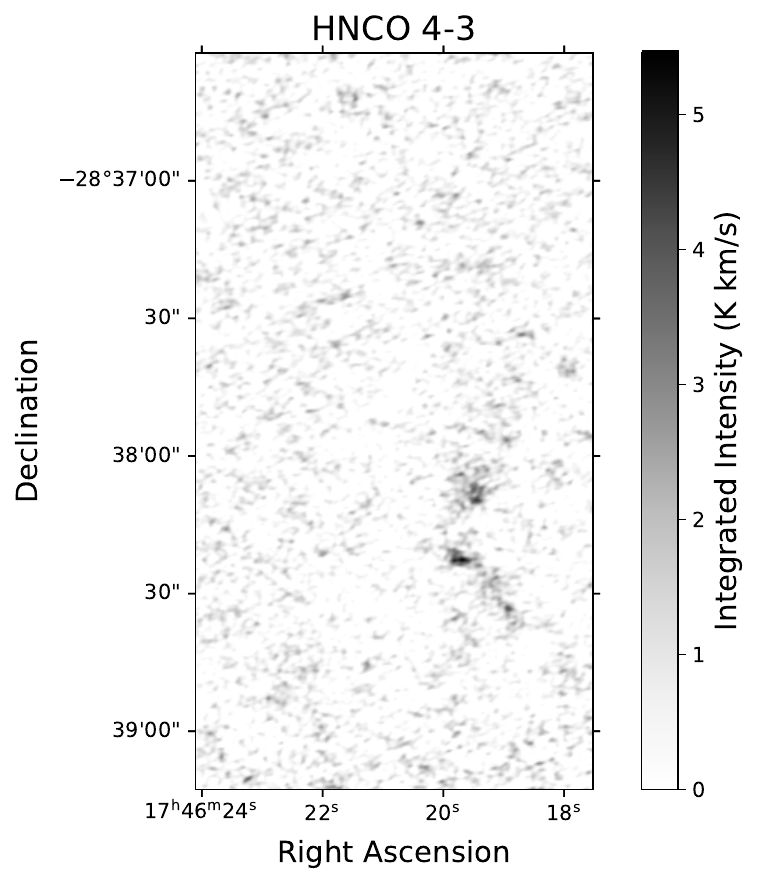}
    \includegraphics[width=0.33\textwidth]{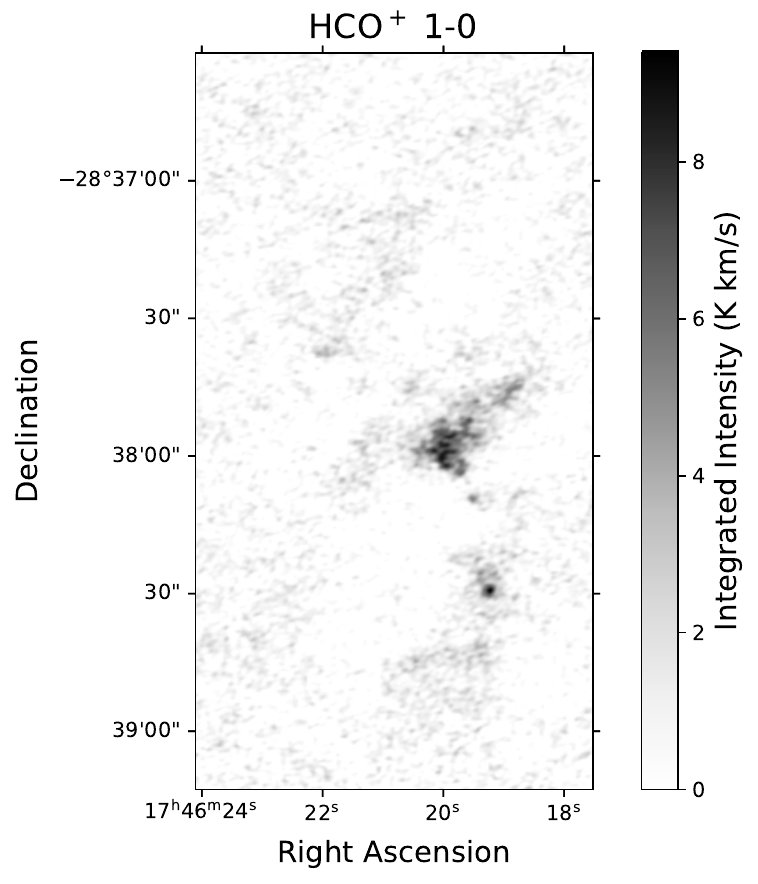}
    \includegraphics[width=0.33\textwidth]{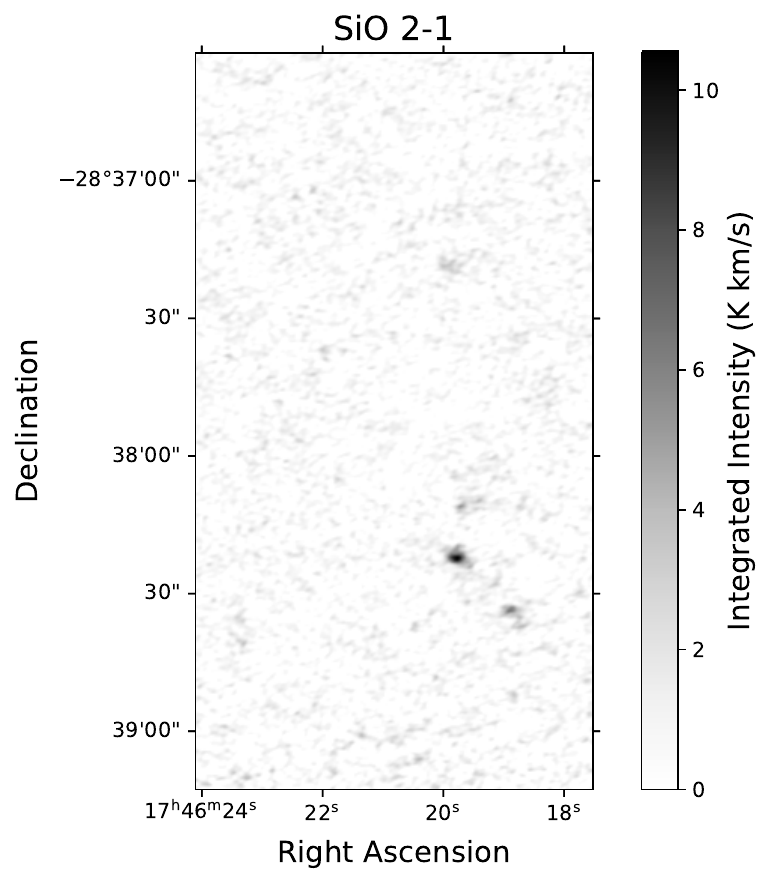}
    \includegraphics[width=0.33\textwidth]{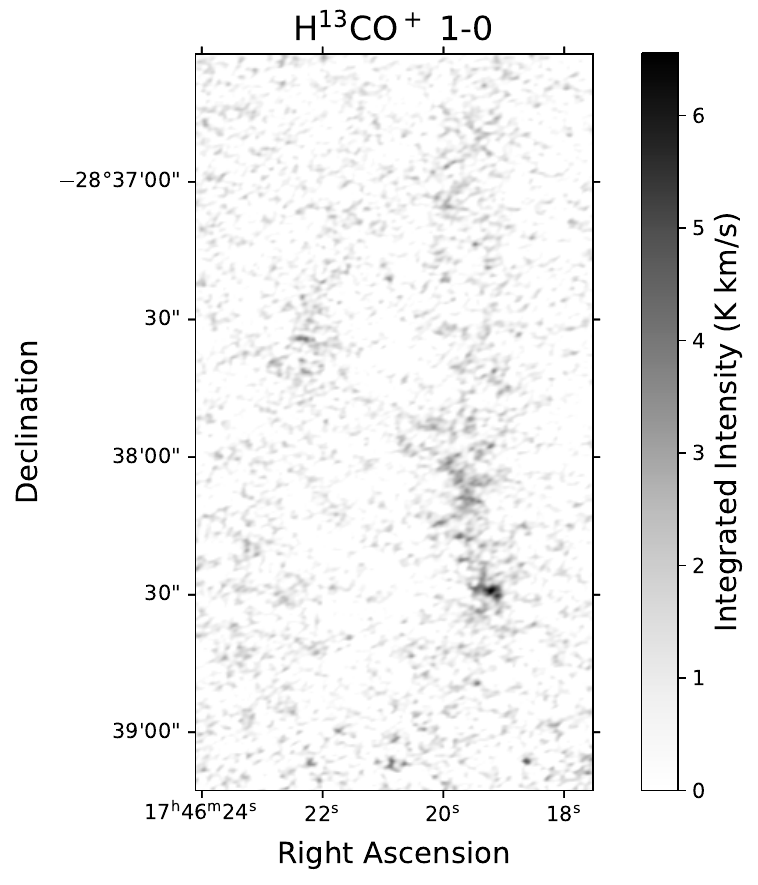}
    \includegraphics[width=0.33\textwidth]{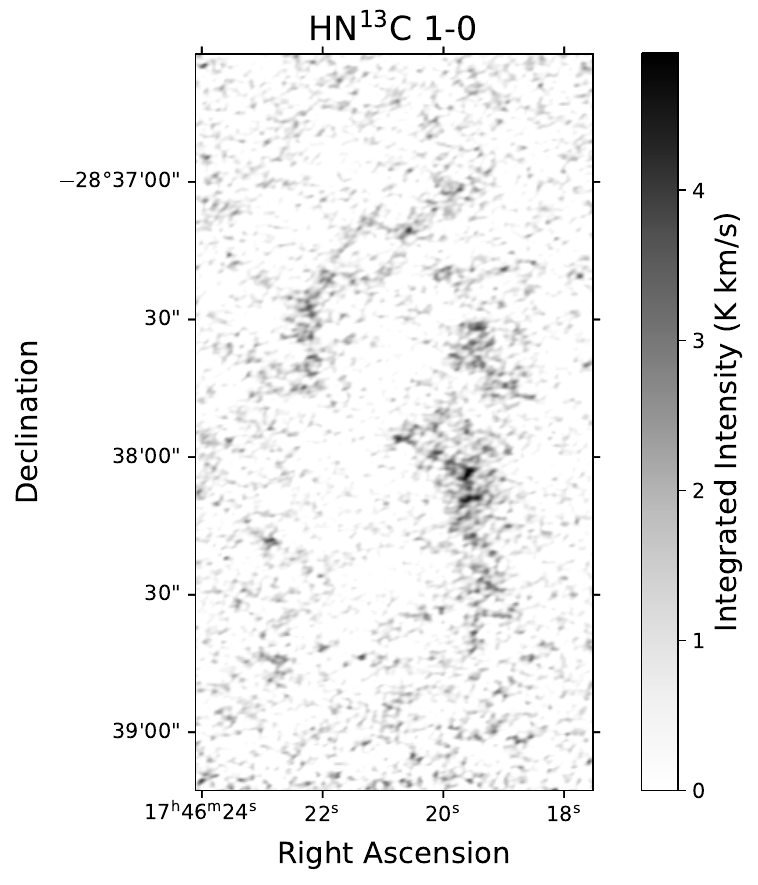}
    \includegraphics[width=0.33\textwidth]{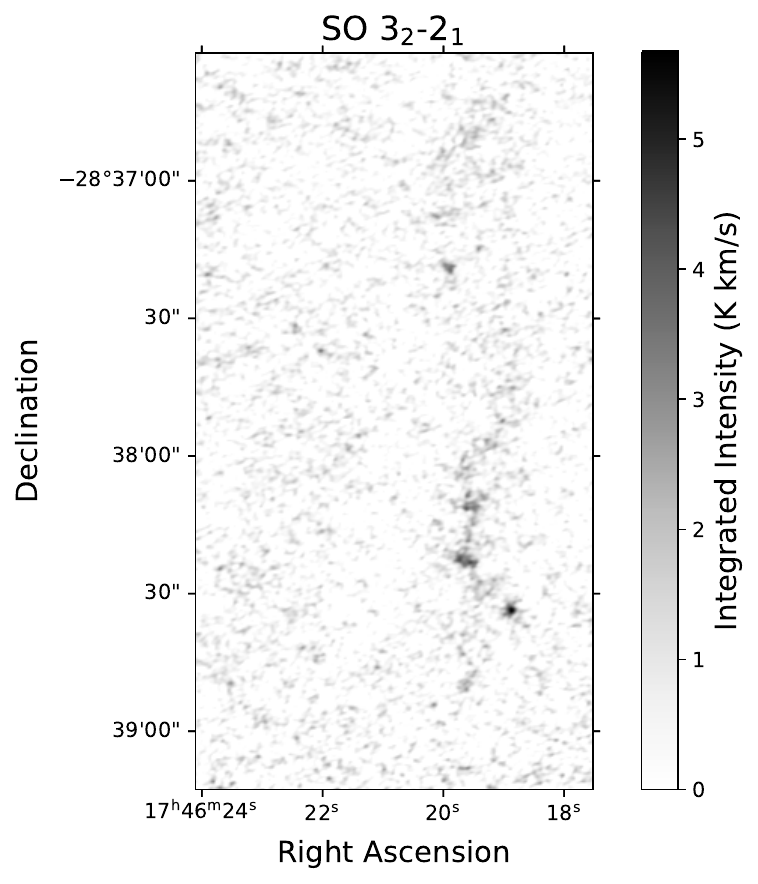}
    \includegraphics[width=0.33\textwidth]{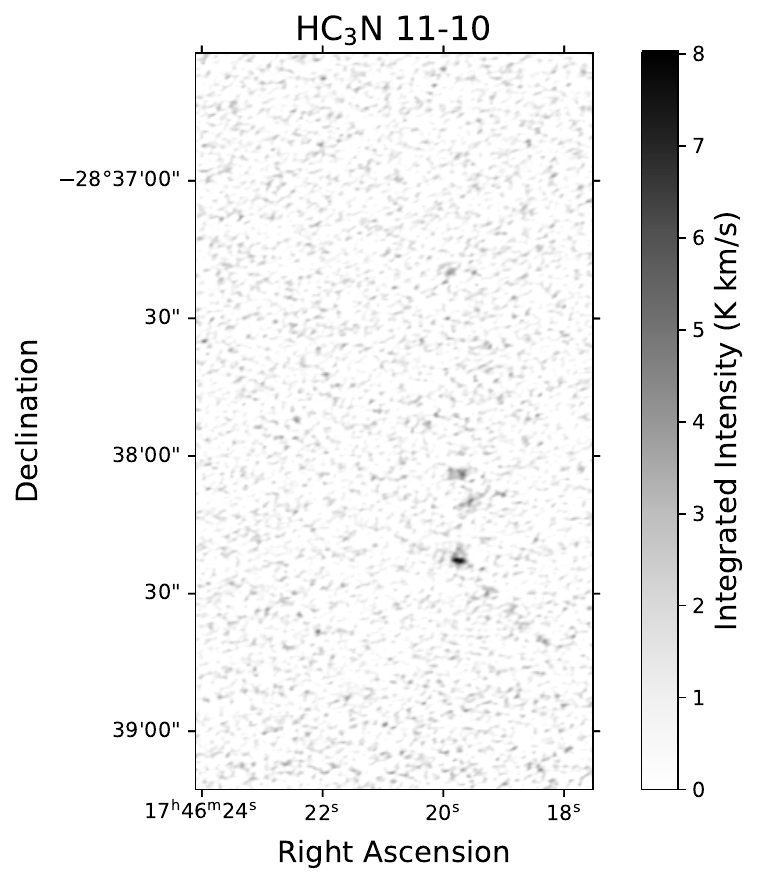}
    \includegraphics[width=0.33\textwidth]{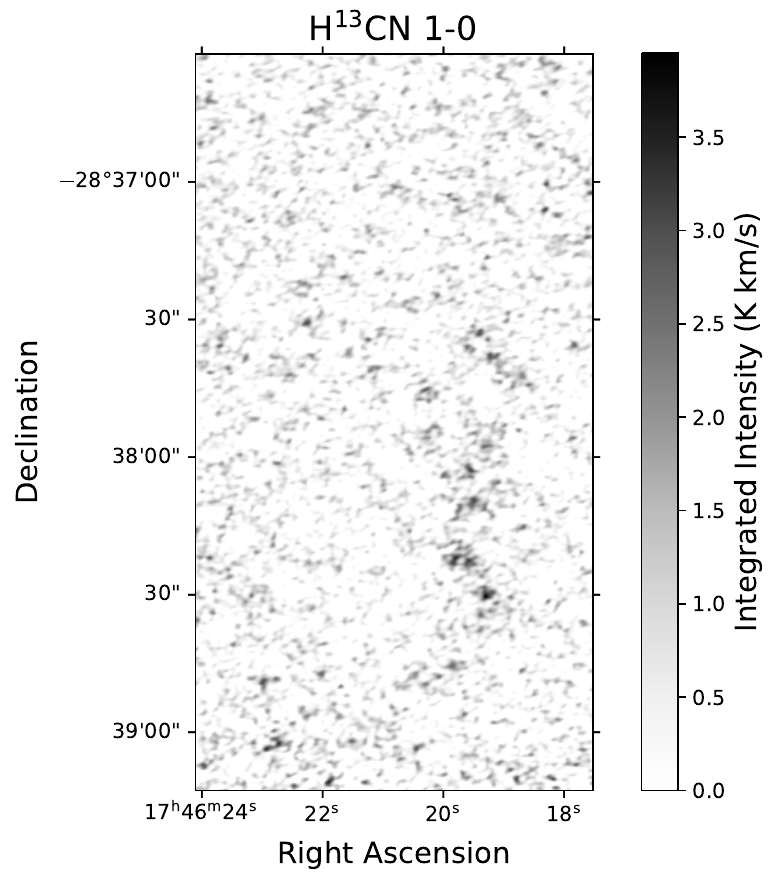}
    \caption{Integrated intensity maps of the filament over the velocities $-56~\kms$ to $-52~\kms$ \referee{for each line detected toward the filament as listed in Table \ref{tbl:lines}}. The most prominent features are the outflows and emission associated with the cores.}
    \label{fig:app-mom0}
\end{figure*}

Figure \ref{fig:app-mom0} shows velocity integrated intensity maps of all of the molecular lines with notable detections for the filament and/or its outflows in the ACES bands. Note that some lines are only detected in the outflows. 

\section{Attempted Distance Measurement} \label{app:dist}

We attempted to measure the distance to the filament. The line of sight velocity of the filament, $-55~\kms$, matches the velocity of $-53~\kms$ for the $\mathrm{3~kpc}$ arm at the Galactic Longitude of the Galactic Center \citep{dame2008}. The $\mathrm{3~kpc}$ arm is assumed to be a ring of material at an orbital radius of $\mathrm{3.5~kpc}$ away from the Galactic Center, but exact distances toward material in front of the Galactic Center associated with the arm have not been measured. As the filament is both backlit by the CMZ and associated kinematically with the $\mathrm{3~kpc}$ arm, we attempt to measure the distance to the object. 

First, we try using the TRILEGAL \citep{Girardi2005} stellar population model to estimate the distance. This model simulates a population of stars along a given line of sight. 
We query a $\mathrm{0.1~deg^2}$ field area at Galactic coordinates ($\ell$, b) = (0.4, 0.04), with the JWST NIRCam filters, no dust extinction, and setting the Sun's Galactocentric radius to $\mathrm{8.210~kpc}$ \citep{Gravity2019}. We choose a limiting magnitude of 23 in the F182M filter to remove stars outside of our medium band magnitude limit, and set the distance modulus resolution to $\mathrm{0.05~mag}$. We used the distance modulus given by the TRILEGAL model to apply the \textbf{CT06} extinction law to the magnitudes of each star. 
We then compare the TRILEGAL output to our catalog of stars. 

\begin{figure}
    \centering
    \includegraphics[width=0.5\linewidth]{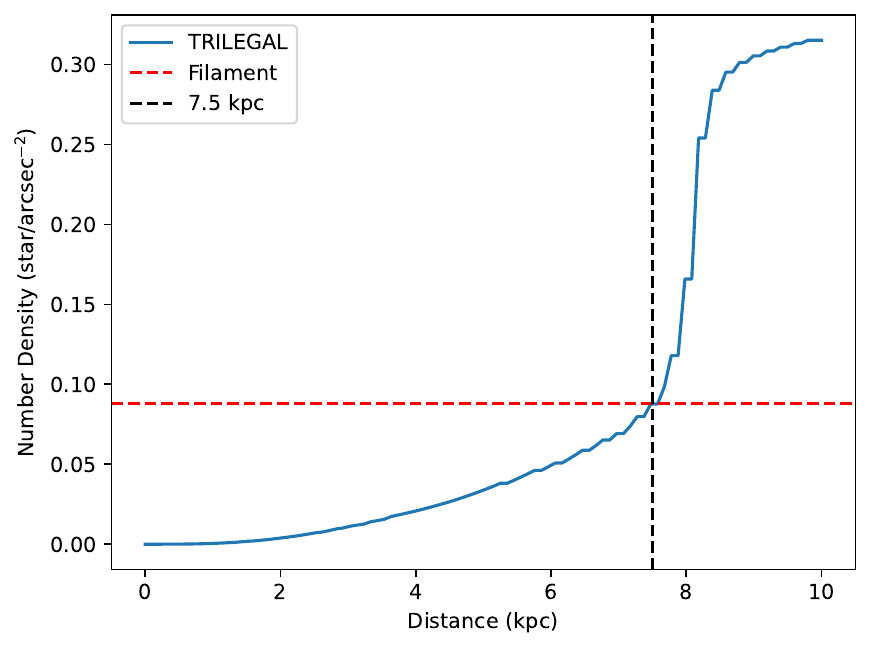}
    \caption{The cumulative number density versus distance for the TRILEGAL model \citep{Girardi2005}. The number density of stars in front of the filament and the distance where that line intersects with the TRILEGAL model are also shown.}
    \label{fig:trilegal}
\end{figure}

To measure the distance to the filament using TRILEGAL, we use a comparison of stellar densities. As distance increases, so does extinction and cumulative stellar density. If we find the stellar density of stars in front of the filament and compare to TRILEGAL, we could find the distance to the filament. 
To find the stars in front of the filament, we limit our catalog to stars in the area of the darkest part of the filament. Using the percentile filter image for F182M from \ref{fig:app-perfilt}, we cut out the area of the filament and then create a mask where the image has values $< 1.25$. 
As the percentile filter images are smoothing over all emission in the field, this selects for the area in the image with less starlight. This part of the image has an area of $\mathrm{1341.6~arcsec^{-2}}$.
We then select for stars with colors of [F182M]-[F410M]$<2$ and magnitudes $15 < [\mathrm{F182M}] < 20$, where the color selects for when stars begin to drop out or become reddened due to the filament, to select for stars in from of the filament. Dividing the number of stars in front of the filament by the area, we measure a stellar density of $\mathrm{0.088~arcsec^{-2}}$.
Plotting the measured stellar density versus a cumulative stellar density versus distance plot using the TRILEGAL data, we find a distance of $\mathrm{7.5~kpc}$. 

Given that a distance of $\mathrm{7.5~kpc}$ is quite close to the Galactic Center, we examine the model and data further to make sure that we were calibrating the TRILEGAL data correctly. What we find is that the TRILEGAL data predict a much smaller stellar density toward the Galactic Center than measured with JWST. The number density of stars in the TRILEGAL model levels off at a smaller stellar density than the JWST catalog, meaning that there are many more stars in the CMZ than predicted by TRILEGAL. Thus, we find that these results are not trustworthy, and the distance from this method is an upper limit. 

Next, we examine using optical surveys with distances to stars to find a lower limit on the distance. First, we query the Gaia survey \citep{Gaia2016, Gaia2023} for stars with distances \citep{Lindegren2021, Gaia2021} within our field using \texttt{astroquery} \citep{astroquery2019}. A total of 565 Gaia sources are within our field of view, which we then matched with our JWST catalog for a total of 410 matches. Of these, only 87 had associated Gaia distance measurements, and they were all at distances $<$ $\mathrm{1~kpc}$, likely due to the amount of extinction in the direction of the Galactic Center in optical wavelengths and Gaia's sensitivity limits. There are too few stars in our field of view from Gaia to make a distance estimate, and there were no stars behind or within the filament in this catalog.

We also attempted to use another survey to estimate the distance, the DECaPS2 (Decaps) survey \citep{Schlafly2018, Saydjari2023}. The Decaps survey goes deeper and includes redder bands than Gaia. We crossmatched the Decaps catalog to our JWST catalog, finding 103 stars with distances matched to our catalog. We used the 50th percentile of the posterior distribution of fitted distances. While these distances range from less than $\mathrm{1~kpc}$ to over $\mathrm{12~kpc}$, easily encompassing the range expected for the $\mathrm{3~kpc}$ arm filament, the distances are not very well matched to the extinction-dominated color observed with JWST. Decaps stars with longer distances match with stars with bluer [F182M]-[F410M] colors than stars with shorter distances, making the distance estimate impossible due to a nonsensical distance and extinction relationship. 

\pagebreak
\bibliographystyle{abbrvnat} % openjournal
\bibliography{sources.bib}

\end{document}